\documentclass[twocolumn]{aa}
\usepackage{graphicx}
\usepackage{txfonts}
\usepackage{amsmath}
\usepackage{leftidx}
\usepackage{xcolor}
\usepackage{diagbox}
\usepackage{ulem}

\def\aap{Astron Astroph}

\newcommand{\old}[1]{}

\def\apjs{ApJS}

\def\aap{Astron Astroph}
\def\apjl{ApJ}

\DeclareUnicodeCharacter{2212}{-} 

\begin{document}

\title{Shadow of quadrupole-deformed compact objects in a local dark matter shell}

 \author{Shokoufe Faraji
          \inst{1}
          \inst{2}
        \inst{3}
        \inst{4}
       \and
        Jo\~{a}o Lu\'{i}s Rosa
        \inst{5}
        \inst{6}
          }

  \institute{University of Bremen, Center of Applied Space Technology and Microgravity (ZARM), 28359 Germany
  \and
  Waterloo Centre for Astrophysics, University of Waterloo, Waterloo, Ontario N2L 3G1, Canada
    \and 
    Department of Physics and Astronomy, University of Waterloo, 200 University Avenue West, N2L 3G1 Waterloo, Canada
    \and 
    Perimeter Institute for Theoretical Physics, 31 Caroline Street North, Waterloo, Canada
    \\
  \email{s3faraji@uwaterloo.ca}
         \and
             Institute of Theoretical Physics and Astrophysics, University of Gda\'{n}sk, Jana Ba\.{z}y\'{n}skiego 8, 80-309 Gda\'{n}sk, Poland
             \and
             Institute of Physics, University of Tartu, W. Ostwaldi 1, 50411 Tartu, Estonia\\
             \email{joaoluis92@gmail.com}
                       }

 \date{Received ......; accepted .....}

\abstract
{This work investigates observational properties, namely the shadow and photon ring structure, of emission profiles originating near compact objects. In particular, we consider a distorted and deformed compact object characterised by two quadrupole parameters and surrounded by an optically thin and geometrically thin accretion disk with different emission profiles modelled by Johnson's Standard-Unbound (SU) distribution in the reference frame of the emitter. Under these assumptions, we produce the observed intensity profiles and shadow images for a face-on observer. Our results indicate that, due to the fact that modifications of the quadrupole parameters affect the radius of the innermost stable circular orbit (ISCO) and the unstable photon orbits on the equatorial plane, the observed shadow images and their properties are significantly influenced by the quadrupole parameters and emission profiles. Furthermore, we analyse the impact of the presence of a dark matter halo in the observational imprints considered and verify that both the increase in the matter contained in the halo or the decrease in the length-scale of the halo lead to an increase in the size of the observed shadow. Our results indicate potential degeneracies between the observational features of distorted and deformed compact objects with those of spherically symmetric blackholes, which could be assessed by a comparison with the current and future generation of optical experiments in gravitational physics.}


\maketitle

\section{Introduction}

The Event Horizon Telescope (EHT) Collaboration recently reported 1.3 mm Very Long Baseline Interferometry (VLBI) observations of the centre of both the M87* galaxy and the object Sgr A* in the core of our own galaxy. The EHT Collaboration has reported the observation of an annular feature centred on M87* with a typical radius of $\sim 21 \mu as$. The brightness along the ring is asymmetric mostly because of the Doppler boosting from the matter orbiting around the jet axis, which is inclined by 17 degrees relative to the line of sight \citep{2019ApJ...875L...1E,2019ApJ...875L...2E, 2019ApJ...875L...3E, 2019ApJ...875L...4E,2019ApJ...875L...5E,2019ApJ...875L...6E,2021ApJ...910L..12E, 2021ApJ...910L..13E}. The ring diameter of the central source of our Milky Way, Sgr A*, located at a distance of $8 kpc$ is reported as $\sim 50 \mu as$, with an image that is dominated by a bright, thick ring with a diameter of $51.8 \pm 2.3 \mu as$. This ring has a modest azimuthal brightness asymmetry and a comparatively dim interior. Preferably the dominant model of the accretion disk that reproduces these observations is a magnetically arrested disk with inclination $i\leq 30$ degrees. \citep{2022ApJ...930L..12E,2022ApJ...930L..13E,2022ApJ...930L..14E,2022ApJ...930L..15E,2022ApJ...930L..16E}.

The shadow and surrounding photon ring involve a special curve known as the apparent boundary \citep{1973blho.conf..215B}, and it is well known as the critical curve on the image plane. A light ray with an impact parameter equal to that of the critical curve asymptotically approaches a bound photon orbit when traced backwards from its observation by a distant observer. The existence of such a critical curve produces an observable photon ring: a bright and narrow ring-shaped feature near the critical curve due to the emission of radiation by the optically thin matter from the region the central object, whose photons travel close to the unstable photon orbit before reaching the observer
\citep{1997A&A...326..419J,Falcke_2000,2010ApJ...718..446J}.

Additionally, the light rays associated with the photon ring typically orbit the central object multiple times before reaching the observer. This can lead to a significant increase in brightness that may continue to grow without limit in the absence of absorption, as the optical path lengths near the unstable photon orbit become increasingly long; nevertheless, the specific form of the visible photon ring is not influenced by the astronomical source profile \citep{PhysRevD.102.124004}. Therefore, it can be used as a probe to test general relativity and models for compact objects. In the absence of rotation, one obtains a single unstable circular photon orbit corresponding to an extremum of the effective potential, which is called a photon sphere. When spherical symmetry is lost, the set of unstable photon orbits is deformed and thickened, giving rise to a photon shell \citep{1973grav.book.....M}.

Separating the contributions of the background geometry and the influence of the surrounding flow in shadow images has become a critical challenge in the field. In the case of a spherical accretion configuration, the size of the observed shadow for a Schwarzschild black hole is closely influenced only by the spacetime geometry, without being affected by the emission profile \citep{2019ApJ...885L..33N}. However, considering an optically and geometrically thin disk around a compact object, later studies indicate that the black hole image is composed of a dark region and a bright region consisting of several components: direct emission, lensing rings, and photon rings, and the size of the shadow depends on the location of the inner edge of the accretion disk, and the intensity of the observed luminosity \citep{PhysRevD.100.024018}. Those important features may be regarded as a way to distinguish between black holes and black hole mimickers in the universe \citep{2023Univ....9...88M,2004CQGra..21.1135V,2024PhRvD.109b4042C} 
In fact, the brightness and width of the lensing ring depend on the background geometry and the emission features of the accretion disk surrounding the compact object, and the observed image is dominated by the direct component \citep{Rosa:2022tfv}. Nonetheless, this is only true for axial observations, as the direct and lensed contributions lose their circular structure for inclined observations \citep{Rosa:2023qcv,Rosa:2023hfm,Rosa:2024bqv}. 

If the spacetime under consideration features an innermost stable circular orbit (ISCO), i.e., a radius below which circular orbits become unstable, this radius accurately predicts where the inner edge of the accretion disk lies \citep{2003ApJ...592..354A}. However, in models of optically thin and geometrically thick accretion disks, for which the inner edge of the disk can penetrate inside the unstable photon orbit, then the minimum size of the shadow is not necessarily limited to the critical curve and can be strongly reduced \citep{2024JCAP...05..032Z,Vincent:2022fwj}.

One of the main advantages of shadow observations is to provide a framework with which to probe deviations with respect to the Kerr or Schwarzschild solutions. However, the limited precision of the first set of observations does not allow for a precise identification of the underlying geometry \citep{Wielgus:2021peu,Daas:2022iid,Lin:2022ksb}. Therefore, it is difficult to underestimate theoretical efforts to calculate the forms and properties of shadows produced by black holes or other alternative compact objects in astrophysical environments. Besides these two well-studied Schwarzschild and Kerr hypotheses, at the theoretical level, several alternative models for compact objects have been proposed, featuring a wide range of properties that could potentially fill the gap in the interpretation and understanding of the observational data \citep{2019LRR....22....4C}.

In this work, we are interested in studying the shadow properties in the background of a distorted and deformed compact object due to quadrupole moments. This solution is a generalisation of the $\rm q$-metric previously analysed in other publications, e.g. \citep{osti_4201189,2022Univ....8..195F}. We show that moderate changes in the quadrupole moment could lead to significant alterations in the properties of the shadow.

On one hand, we see that the radius of the photon ring, the innermost stable circular orbit (ISCO), and optical appearance can be different depending on the parameters of the geometry and intensity profile. On the other hand, we can deduce information about the source from its observed intensity profile. The latter has a crucial importance in the test of the nature of compact objects and, therefore, in the test of the strong field regime in the general theory of relativity.

Another important issue in astrophysics is the observation of abnormally high velocities in the outer region of galaxies due to the surrounding invisible dark matter \citep{1993A&A...275...67B,2001ARA&A..39..137S}. The gravitational tidal forces induced by dark matter influence the propagation of light. Therefore, the presence of dark matter can be analysed through the properties of the shadow via the assumption of different equations of state, e.g. \citep{2018JCAP...12..040H,2018JCAP...07..015H,2019PhRvD..99d4015H}. However, in these studies the results are model-dependent. To explore the effect of dark matter, we consider its behaviour as an effective mass, which leads to modifications in the background geometry, e.g. \citep{1997PhRvL..78.2894L,2011RvMP...83..793K,2019PhRvD..99b4007K,2019PhLB..795....1K}.

The paper is organized as follows: in Section \ref{spacetime}, we introduce the generalized $\rm q$-metric; in Section \ref{shadowring} we review the geodesic motion of timelike and null particles in this background, we introduce the accretion disk models, and we produce the observed shadow images and intensity profiles; in Section \ref{darkmatter} we introduce the dark matter model as a modification of the radial mass function and obtain the corresponding observed shadow images and intensity profiles; and in Section \ref{summary} we summarize our results and trace prospects for future work. Throughout the paper, we use a system of geometric units in which $G = 1 = c$ unless stated otherwise.

\section{Generalized $\rm q$-metric spacetime} \label{spacetime}

\begin{figure*}
    \centering
    \includegraphics[scale=0.8]{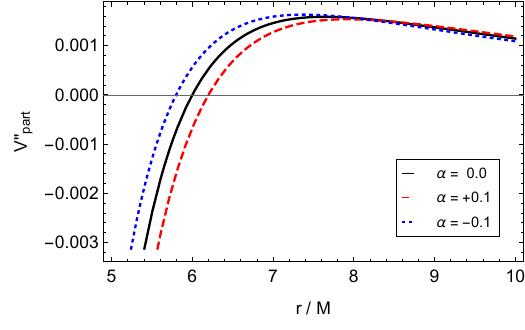}\qquad
    \includegraphics[scale=0.8]{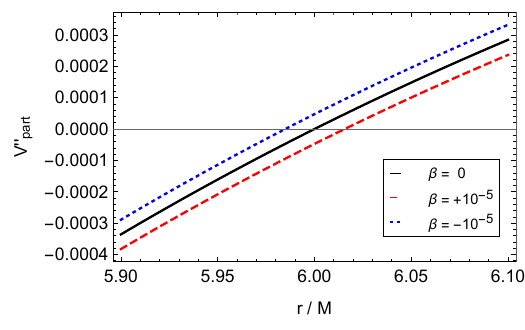}\\
    \includegraphics[scale=0.8]{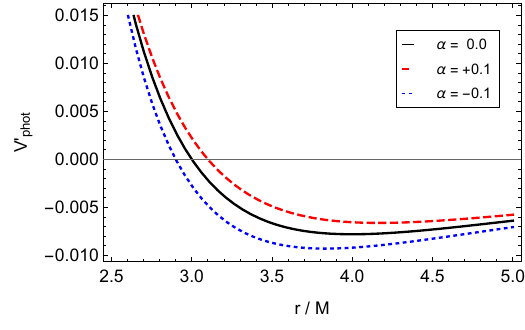}\qquad
    \includegraphics[scale=0.8]{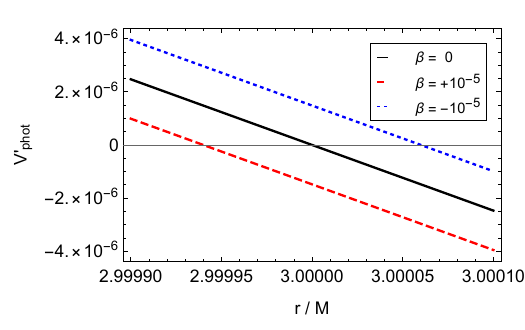}
    \caption{Second radial derivative of the particle effective potential $V''_{\rm part}$ (top row), and first radial derivative of the photon potential $V'_{\rm phot}$, both as a function of the radius $r$, with $\beta=0$ and varying $\alpha$ (left panels) and with $\alpha=0$ and varying $\beta$ (right panels), at the equatorial plane $\theta=\frac{\pi}{2}$. The radius of the ISCO is defined as the radius at which $V''_{\rm part}=0$, and the radius of the photon orbit is defined as the radius at which $V'_{\rm phot}=0$. We observe that a change in $\alpha$ causes the radii of the ISCO and the photon orbit to change in the same direction, whereas a change in $\beta$ causes these radii to change in opposite directions.}
    \label{fig:potentials}
\end{figure*}

One of the generalisations of the Schwarzschild spacetime recurring to the Weyl's family of solutions \citep{doi:10.1002/andp.19173591804} is the $\rm q$-metric. This asymptotically flat solution of the Einstein field equations represents the exterior gravitational field of an isolated static and axisymmetric mass distribution containing a quadrupole moment, and it can be used to investigate the exterior fields of slightly deformed astrophysical objects in the strong-field regime \citep{Quevedo:2010vx}. This metric has been further generalised by relaxing the assumption of an isolated compact object, or equivalently asymptotic flatness, by considering the presence of an exterior distribution of mass in its vicinity \citep{2022Univ....8..195F}. Since this spacetime is not asymptotically flat by construction, the solution presented here is a vacuum solution of the Einstein field equations in the region near the central deformed object. The extension of this vacuum region depends on the chosen values of the parameters—smaller parameter values lead to a more extended vacuum region. Beyond this vacuum region, the spacetime could be extended by introducing a proper stress-energy tensor, which can be interpreted as a general external distribution of matter. Such a generalisation resembles a magnetic surrounding \citep{1976JMP....17...54E} and it is also characterised by multipole moments. However, due to the subdominant effects of the higher multipole moments, it is appropriate to consider only the contributions to the metric up to the quadrupole moment for analyses in an astrophysical context. Therefore, in this spacetime, there are two independent quadrupole moments: one responsible for the deformation of the central object, $\alpha$, and the other associated with the relaxation of asymptotic flatness, linked to the concept of an external matter distribution, $\beta$. Here we adopt a set of coordinates analogous to the usual spherical coordinates $\left(t,r,\theta,\phi\right)$ in spherically symmetric spacetimes\footnote{In this spacetime, the coordinate $r$ represents a radial parameter where $r=\text{const.}$ hypersurfaces are topologically spheres. However, due to the deformation of spacetime, the geometric properties of these hypersurfaces, such as surface area and curvature, are altered. This modification is reflected in the metric components, which adjust the standard interpretation of distances and angles in spherical coordinates.}. the line element describing this generalized $\rm q$-metric is written as
\begin{align}\label{eq:metric}
   ds^2 = &\left(1-\frac{2M}{r}\right)^{1+\alpha} e^{2\hat{\psi}} dt^2 - \left(1-\frac{2M}{r}\right)^
    {-\alpha} e^{-2\hat{\psi}}\nonumber\\
    &\left[ \left(1+\frac{M^2\sin^2\theta}{r^2-2Mr}\right)^{-\alpha(2+\alpha)}e^{2\hat{\gamma}}\left(\frac{dr^2}{1-\frac{2M}{r}}+r^2d\theta^2\right)\right. \nonumber\\
  &\left.+r^2\sin^2\theta d\phi^2\right],\
\end{align}
where $M$ is a parameter that plays the role of the mass of the central object, and the metric functions $\hat\psi$ and $\hat\gamma$ that control the distortion of the spacetime also up to quadrupole read as follows
 
\begin{align}\label{gammapsi}
	\hat{\psi} & = \frac {\beta}{2}\left[\left(\frac{r}{M}-1\right)^2\left(3 \cos^2 \theta-1\right) +\sin^2 \theta \right], \\
	\hat{\gamma} & = \beta \sin^2\theta \left[-2  (1+\alpha)\left(\frac{r}{M}-1\right)\right.\nonumber\\
 &\left.+\frac{\beta r}{4M}\left(\frac{r}{M}-2\right)\left((-9\cos^2\theta+1)\left(\frac{r}{M}-1 \right)^2 -\sin^2\theta \right)\right]\nonumber \,.
\end{align}
where $\alpha$ and $\beta$ are quadrupole parameters. The coordinates range in the intervals $t \in (-\infty, +\infty)$, $r \in (0, +\infty)$, $\theta \in [0,\pi]$, and $\phi \in [0, 2\pi)$. Under the assumptions outlined above, this metric contains three free parameters: the deformation parameter $\alpha$, the distortion parameter $\beta$, and the total mass $M$ that can be absorbed through a redefinition of the radial coordinate. In the equatorial plane, described by $\theta=\pi/2$, the metric functions reduce to
\begin{align}\label{gammapsieq}
	\hat{\psi} & = \frac {\beta}{2}\left[-\left(\frac{r}{M}-1\right)^2+1\right],\\
	\hat{\gamma} & = \beta \left[-2  (1+\alpha)\left(\frac{r}{M}-1\right)+\frac{\beta r}{4M}\left(\frac{r}{M}-2\right)\left(\left(\frac{r}{M}-1 \right)^2 -1 \right)\right]\nonumber \,.
\end{align}
In fact, the presence of a quadrupole can have a notable impact on the geometric properties of space-time when compared to the Schwarzschild solution, see e.g. \citep{2021A&A...654A.100F,10.1093/mnras/stac882}. Mathematically, the deformation parameter is bounded by $\alpha \in (-1,\infty)$. However, in an appropriate astrophysical context, other studies see e.g. \citep{2021A&A...654A.100F,10.1093/mnras/stac882}, have demonstrated that the most relevant domain for $\alpha$ is almost $(-1,1)$. It is important to note that geodesics are influenced by both parameters $\alpha$ and $\beta$ simultaneously. In fact, calculating the radius of the innermost stable circular orbit (ISCO) and of the unstable photon orbit, which is done by setting the second and first derivatives of the respective effective potentials to zero, leads to a relationship between $\alpha$ and $\beta$. This happens because, for a given value of $\alpha$ the range of the parameter $\beta$, specifically in the equatorial plane, depends on $\alpha$ and arises directly from solving the mentioned equations. In addition, the range of the parameter $\beta$ turns out to be significantly smaller than the latter. Additionally, in this work, we are interested in configurations that are approximately spherically symmetric, and consequently, the quadrupole parameters should be relatively small in magnitude. To guarantee that this approximation of spherical symmetry is valid, one needs to keep the absolute values of the parameters $\alpha$ and $\beta$ below the values $0.01$ and $10^{-6}$, respectively, which guarantees that the deviations from spherical symmetry of the $g_{tt}$, $g_{rr}$, and $g_{\phi\phi}$ components of the spacetime metric are always below $1\%$ in the region where photons propagate (see Appendix \ref{app:approx}).
This solution is the exact superposition of an oblate or prolate compact object, surrounded by an external distribution of matter. Recently, the impact of superposed fields in the motion of free test particles in this spacetime was studied \citep{2022Univ....8..195F}, and it was shown how the positions of important astrophysical orbits, such as ISCO for particles, and photon circular orbits, depend on the parameters of the metric. In Figure \ref{fig:potentials}, we plot the second derivative of the particle potential $V_{\rm part}=-g_{tt}\left(\frac{L^2}{r^2}+1\right)$, where $L$ is the angular momentum of the particle, and the first derivative of the photon effective potential $V_{\rm phot}=-\frac{g_{tt}}{r^2}$. The results indicate that an increase in $\alpha$ shifts the inflection points of the particle effective potential and the extrema of the photon effective potential to larger radial coordinates. In contrast, an increase in $\beta$ causes the behaviour of these two features to shift in opposite directions due to the effect of external distribution. In this case, the unstable photon orbit, being closer to the object than ISCO, feels the gravitational pull of the central object more strongly than the influence of the external fields. For instance, if $\beta>0$ the external quadrupole tends to push the unstable photon orbit inward, increasing the curvature near the central object. The ISCO, located farther out, feels a weaker pull from the central object but a stronger influence from the external quadrupole, which causes it to shift outward.

Moreover, from an astrophysical perspective, the negative values of the quadrupole parameters are rather counter-intuitive since they describe prolate compact objects, while positive values correspond to oblate compact objects that are widely observed in the universe.

\section{Shadows and light rings} \label{shadowring}

The black hole shadow for a far-away observer is defined as the set of directions in the sky, propagated backwards in time, that cross the event horizon. Its boundary consists of those rays that asymptotically approach the surfaces of the underlying spacetime on which photons undergo unstable closed orbits, i.e., the photon shell (PS) or critical curves, with the same specific energy and angular momentum that are conserved quantities. In fact, the shape and size of the shadow depend on the geometry of the background spacetime, which requires the analysis of the light-ray or null geodesics of a test particle.

\subsection{Null geodesics and impact parameter}

The Lagrangian $\mathcal{L}= \frac{1}{2} g_{\mu \nu}\dot{x}^\mu \dot{x}^\nu$ describing the motion of a particle, in a spacetime described by a diagonal metric in spherical coordinates and restricted to the equatorial plane reads as

\begin{align} \label{lagrang}
    \mathcal{L}=  \frac{1}{2} \left[ g_{tt}\dot{t}^2- g_{rr}\dot{r}^2 - g_{\phi \phi}\dot{\phi}^2\right]
\end{align}
where $\dot{x}^\mu\equiv\frac{dx^\mu}{d\tau}$ and $\tau$ is proper time. In addition, there exist two conserved quantities associated with time translation and angular momentum, which are defined as

\begin{align} \label{el}
    E=\frac{\partial \mathcal L}{\partial \dot{t}} = g_{tt}\dot{t}, \quad \quad  L=\frac{\partial \mathcal L}{\partial \dot{\phi}} = -g_{\phi \phi}\dot{\phi}.
\end{align}
Solving Eq. \eqref{el} with respect to $\dot t$ and $\dot \phi$ and inserting the result into Eq. \eqref{lagrang} under the restriction $g_{\mu \nu}\dot{x}^\mu \dot{x}^\nu =0$ for null geodesics, one obtains

\begin{align} \label{rdot}
    \dot{r}^2 = \frac{1}{g_{rr}}\left[\frac{E^2}{g_{tt}}-\frac{L^2}{g_{\phi \phi}}\right].
\end{align}
For a given photon propagating in the background spacetime, if $\dot r$ vanishes at a radius larger than the unstable photon orbit, this implies that the photon reaches a radius of closest approach and subsequently propagates back to infinity. If this situation is never verified, then the photon is captured by the central object. Introducing the impact parameter $b=\frac{L}{E}$ and rewriting Equation \eqref{rdot}, one can obtain the so-called critical impact parameter $b_{\rm crit}$, which discriminates between captured and uncaptured photons. This can be done by solving the equations $\dot r=0$ and $\ddot r=0$ for circular orbits, which leads to

\begin{align}
    &g_{\phi \phi} - g_{tt}b_{\rm crit}^2=0,\\
    &{g}^\prime_{tt} g^2_{\phi \phi} -b_{\rm crit}^2{g}^\prime_{\phi \phi} g^2_{tt}=0,\
\end{align}
where a prime denotes a derivative with respect to $r$, and the result does not depend on $g_{rr}$. For example, in the Schwarzschild case $b_{\rm crit}=3\sqrt 3$. In our setup, as the radius of the PS depends on the quadrupole parameters, different choices imply different values of $b_{\rm crit}$. If the impact parameter $b < b_{\rm crit}$ for a given photon, the photon is captured by the compact object. Instead, for $b=b_{\rm crit}$ the photon revolves infinitely around the central compact object. Finally, for $b > b_{\rm crit}$  the photon is scattered back and may reach the observer. 

In addition, Eq. \eqref{rdot} can be conveniently rewritten in the following form from which one can obtain the trajectories of the lightrays

\begin{align}\label{eq:geodesic}
\frac{d\phi}{dr}=\pm \frac{b}{g_{\phi \phi}}\frac{\sqrt{-g_{tt}g_{rr}}}{\sqrt{1+\frac{b^2 g_{tt}}{g_{\phi \phi}}}}.
\end{align}
The signs ($\pm$) represent ingoing ($-$) and outgoing ($+$) null geodesics. In what follows, we perform a numerical integration of Eq. \eqref{eq:geodesic} using a Mathematica-based ray-tracing code that has been previously used in several publications to simulate the observational properties of compact objects (\citep{Rosa:2022tfv,Rosa:2023qcv,Rosa:2023hfm,Olmo:2023lil,Guerrero:2022msp,Guerrero:2022qkh,Olmo:2021piq,Guerrero:2021ues,PhysRevD.109.064027,PhysRevD.109.084002,2024JCAP...07..046M}). Note that a given ingoing geodesic transitions into an outgoing geodesic at the radius of the closest approach, i.e., when $d\phi/dr$ diverges, at which point one should invert the sign of Eq. \eqref{eq:geodesic} to proceed with the integration. Note that, contrary to what happens in spherically symmetric spacetimes, the right-hand side of Eq. \eqref{eq:geodesic} depends explicitly on complicated functions of $\theta$, which is a major obstacle in the numerical resolution of the equation. To simplify this procedure and improve the computational time, we employ the spherically symmetric approximation, i.e., we consider small values of the quadrupole parameters and we restrict the analysis to the equatorial plane $\theta=\frac{\pi}{2}$. The errors incurred by taking this approximation are always smaller than $1\%$ for the situations analysed in this work (see Appendix \ref{app:approx} for more details on the validity of this approximation). The study of circular geodesics i.e. $\dot{r}=0$ and $\ddot r=0$, the radius of these circular geodesics is related to $\alpha$ and $\beta$ through the following relation\citep{2022Univ....8..195F}

\begin{align}\label{curveplight1}
\alpha=\beta\frac{r}{M} \left(\frac{r}{M}-1\right) \left(\frac{r}{M}-2\right)+\frac{1}{2}\left(\frac{r}{M}-3\right),
\end{align}
where in the limit $\beta=0$ one recovers the radius of the PS for the $\rm q$-metric $r_{\rm LR}=M(3+2\alpha)$, and in the limit $\alpha=\beta=0$ one recovers the Schwarzschild radius of the PS $r_{\rm LR}=3M$. In addition, one can have one bound unstable photon orbit in the equatorial plane for any given valid $\beta<-\frac{1}{2}$ with

\begin{align}\label{qlight}
\beta = -\frac{1}{2\left[3\left(\frac{r}{M}-1\right)^2-1\right]}.
\end{align}
This is not the case either in the Schwarzschild spacetime, nor in the $\rm q$-metric. In fact, this arises due to the existence of a quadrupole related to the external source and indicating that, for certain negative values of $\beta$, an unstable bound photon orbit exists in the equatorial plane.  It is necessary to exclude these orbits in order to conduct a study on shadows.

\subsection{Accretion flow models and intensity profiles}

The observation of the shadow requires a large enough angular resolution; however, there is a backdrop of sufficiently bright light sources against which the shadow can be seen as a dark spot. For example, the emission of the surrounding matter can partially obscure any shadow. In this section, we consider optically and geometrically thin accretion disks in the equatorial plane surrounding the compact object, to investigate the observed specific intensity. Furthermore, since the optically thin disk is transparent to its own radiation, photons turning more than one half-orbit around the compact object produce a thin photon ring in superposition with direct emission. Consequently, the observed photon ring is broken into an infinite sequence of self-similar photon rings forming several loops and stacking on the direct emission.

We assume that the accretion disk is located in the static rest frame and emits isotropically. More importantly, the whole setup should be close enough to the horizon since photons emitted from this vicinity with a small enough impact parameter are absorbed by the central compact object and cast a shadow. In fact, the inner part of the accretion disc flow has the largest density, therefore, most of the radiation comes from this region with the highest temperature. If the disk has no magnetic field with sub-Eddington luminosities, then, with high accuracy, the place of the innermost stable circular orbit (ISCO) is considered as the inner edge of the thin disk \citep{2003ApJ...592..354A} where most of the luminosity comes from. Therefore, it can be used to infer possible luminosity and intensity profiles for different accretion disk models. However, in what follows, we consider emission from an optically thin and geometrically thin region near the compact object, with three different stationary accretion disk models. In addition, the observer stands on the vertical axis $\theta=0$. Under those considerations, the intensity models are as follows:

\begin{table}
    \centering
    \begin{tabular}{|c|c|c|c|} \hline
    \diagbox{$\beta$}{$\alpha$}  & $-0.01$ & $0$ & $0.01$ \\  \hline 
      $-10^{-6}$  & \begin{tabular}{c}$5.94346M$\\$2.98001M$\end{tabular} & \begin{tabular}{c}$5.99844M$\\$3.00001M$\end{tabular} & \begin{tabular}{c}$6.05336M$\\$3.02001M$\end{tabular} \\ \hline 
      $0$  & \begin{tabular}{c}$5.94497M$\\$2.98000M$\end{tabular}  & \begin{tabular}{c}$6.00000M$\\$3.00000M$\end{tabular} & \begin{tabular}{c}$6.05497M$\\$3.02000M$\end{tabular}  \\ \hline 
      $10^{-6}$  & \begin{tabular}{c}$5.94648M$\\$2.97999M$\end{tabular} & \begin{tabular}{c}$6.00156M$\\$2.99999M$\end{tabular} & \begin{tabular}{c}$6.05659M$\\$3.01999M$\end{tabular} \\ \hline
    \end{tabular}

    \vspace{0.2cm}
    
    \caption{Values for the radii of the ISCO (top value) and the radius of the LR (bottom value) for the different combinations of $\alpha$ and $\beta$ considered in this work, with a precision up to 5 decimal places.}
    \label{tab:radii_values}
\end{table}

\begin{enumerate}[I.]
    \item  The ISCO model: In this model, the luminosity of the accretion disk increases monotonically from $r=\infty$ to $r_{\rm ISCO}$, where it achieves a maximum, and then abruptly decays in the region $r<r_{\rm ISCO}$.

\vspace{0.2cm}
    
    \item The LR model: Even though the circular orbits for $r<r_{\rm ISCO}$ are unstable, one can still argue that unstable orbits exist in the region $r_{\rm LR}<r<r_{\rm ISCO}$. Thus, in this model, it is supposed that the luminosity of the accretion disk increases monotonically from $r=\infty$ to $r_{\rm LR}$, where it peaks, and then abruptly decays in the region $r<r_{\rm LR}$.

\vspace{0.2cm}
    
    \item  The EH model: In this model, the emissivity extends to the event horizon and is not truncated at any larger radius. This model is motivated by the expectation that the region $r<r_{\rm LR}$ is populated by matter falling into the central compact object all the way down to the Event Horizon (EH), which for the Schwarzschild spacetime ($\alpha=\beta=0$) stands at $r_{\rm EH}=2M$. We thus consider an additional intensity model for which the luminosity increases monotonically from $r=\infty$ and peaks at $r_{\rm EH}$. 
\end{enumerate}

\begin{table}
    \centering
    \begin{tabular}{|c|c|c|c|}\hline
         & $\gamma$ & $\mu$ & $\sigma$ \\ \hline
       ISCO & $-2$ & $r_{\rm ISCO}$ & $\frac{1}{4}M$ \\
       LR & $-2$ & $r_{\rm LR}$ & $\frac{1}{8}M$ \\
       EH & $-3$ & $2M$ & $\frac{1}{8}M$ \\\hline
    \end{tabular}

    \vspace{0.2cm}
    
    \caption{Values of the parameters $\gamma$, $\mu$ and $\sigma$ for the three  disk profiles considered in this work using Eq. \eqref{eq:disk_profile}. The radii $r_{\rm ISCO}$ and $r_{\rm LR}$ differ for different combinations of $\alpha$ and $\beta$. The explicit values can be found in Table \ref{tab:radii_values}.}
    \label{tab:disk_parameters}
\end{table}

\begin{figure}
    \centering
    \includegraphics[scale=0.9]{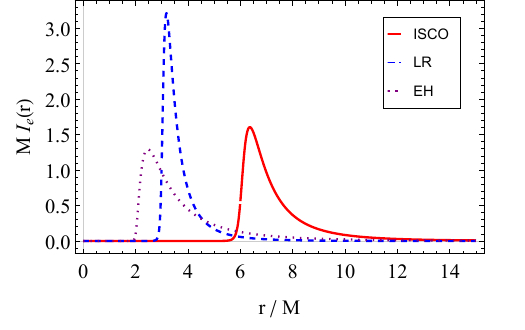}
    \caption{Luminosity profiles for the three models of accretion disks considered: the ISCO model (red solid), the LR model (blue dashed), and the EH model (purple dotted), for $\alpha=0$ and $\beta=0$.}
    \label{fig:disks}
\end{figure}

Note that for $\alpha\neq 0$ the spacetime does not have an event horizon but a naked singularity at $r=2M$. However, for the sake of notation simplicity, we stick to the symbol $r_{\rm EH}=2M$ and pictorial representation for these solutions. To model the three intensity profiles described above, we recur to an emission profile published recently in \citep{PhysRevD.102.124004} and shown to be in close agreement with observational predictions and general relativistic magneto-hydrodynamics (GRMHD) simulations. The intensity profile $I\left(r\right)$ of this model is derived from Johnson’s SU distribution, which is a free function with a smooth profile, mostly concentrated within a few Schwarzschild radii of the central object, and increases as we approach the compact object. This emission profile is given by

\begin{equation}\label{eq:disk_profile}
I\left(r,\gamma,\mu,\sigma\right)=\frac{e^{-\frac{1}{2}\left[\gamma+\text{arcsinh}\left(\frac{r-\mu}{\sigma}\right)\right]^2}}{\sqrt{\left(r-\mu\right)^2+\sigma^2}},
\end{equation}
where $\gamma$, $\mu$, and $\sigma$ are free parameters that shape the behavior of the function $I\left(r\right)$. The parameter $\gamma$ controls the rate of increase of the luminosity from infinity down to the peak; the parameter $\mu$ induces a translation of the profile and controls the radius at which the luminosity peaks; and the parameter $\sigma$ controls the dilation of the profile. The values of the parameters $\gamma$, $\mu$ and $\sigma$ for each of the three disk models mentioned above, i.e., the ISCO, LR and EH disk models, are specified in Table \ref{tab:disk_parameters}, where $r_{LR}$ and $r_{ISCO}$ are provided in Table \ref{tab:radii_values}, and the respective intensity profiles are plotted in Figure  \ref{fig:disks}.  In this Figure, different values of $\alpha$ and $\beta$ are not plotted separately as they correspond solely to a translation of the ISCO and LR models in the horizontal axis. However, we adapted these models in our set-up considering the place of ISCO and light ray is changing with $\alpha$ and $\beta$ \citep{2022Univ....8..195F}.

Note that the intensity profiles plotted in Figure \ref{fig:disks} correspond to the intensity profiles in the reference frame of the accretion disk. The frequencies in the reference frame of the observer and the emitter, are denoted by $\nu_o$ and $\nu_e$, respectively. These two frequencies are related via the gravitational redshift as $\nu_0=\sqrt{g_{tt}}\nu_e$, then the associated specific intensity $i\left(\nu\right)$ scales as 

\begin{align}
 i\left(\nu_o\right)=\left(\nu_0/\nu_e\right)^3i\left(\nu_e\right)=g_{tt}^\frac{3}{2} i\left(\nu_e\right).   
\end{align}
For each time that a light ray intersects the equatorial plane, it gathers additional brightness from the disk emission. Due to the extra brightness contributions, the observed intensity is described as the sum of intensities from each intersection. Consequently, the total intensity $I$ in the reference frame of the observer can be written as 
\begin{align}\label{eq:redshiftdis}
I_o(r)=\sum_{n}g_{tt}^{\frac{3}{2}}\left(r\right) I_{e}\left(r\right)|_{r=r_n(b)},
\end{align}
where $n=\frac{\theta}{2\pi}$, and $r_n(b)$ is the demagnification factor $\frac{dr}{db}$ which represents the intersection radius with the equatorial plane for the nth time on its backwards journey, where $b$ is the impact parameter. The continuous version of Eq. \eqref{eq:redshiftdis} is also taken into consideration in our Mathematica-based code to produce the intensity profiles in the reference frame of the observer.

\subsection{Shadows and observed intensity profiles}

\begin{figure*}
    \centering
    \includegraphics[scale=0.5]{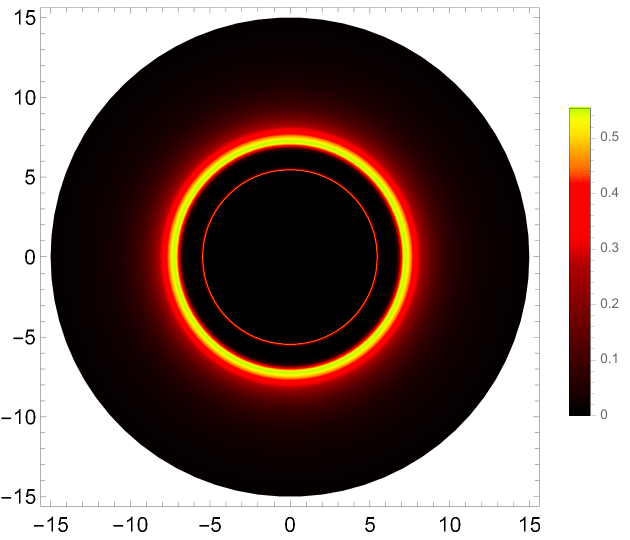}\qquad
    \includegraphics[scale=0.5]{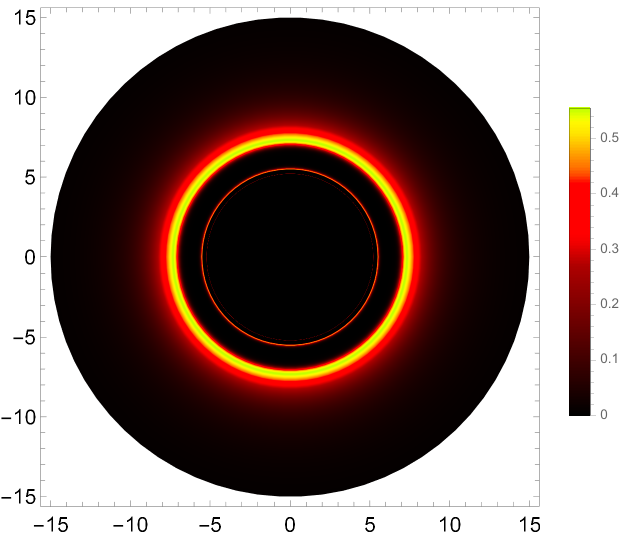} \qquad
    \includegraphics[scale=0.5]{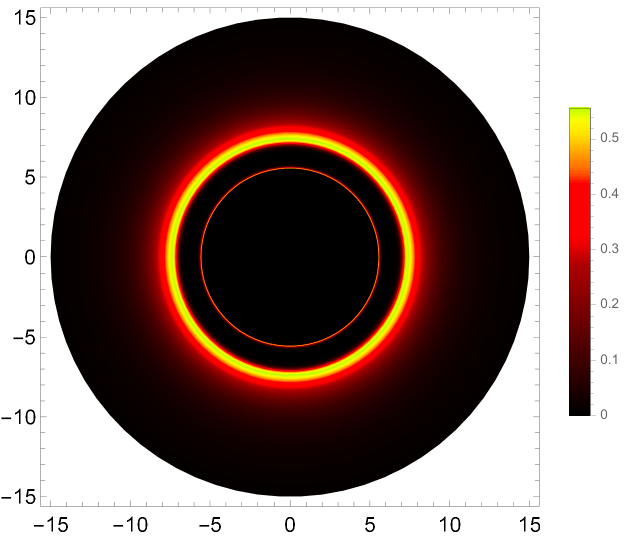}\\
    \includegraphics[scale=0.5]{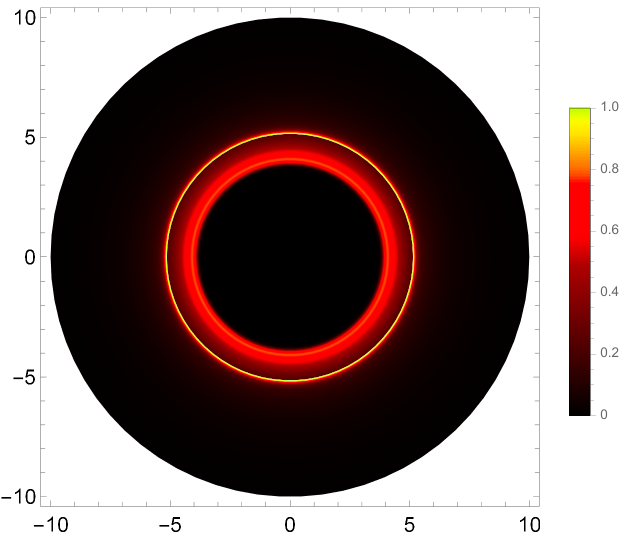}\qquad
    \includegraphics[scale=0.5]{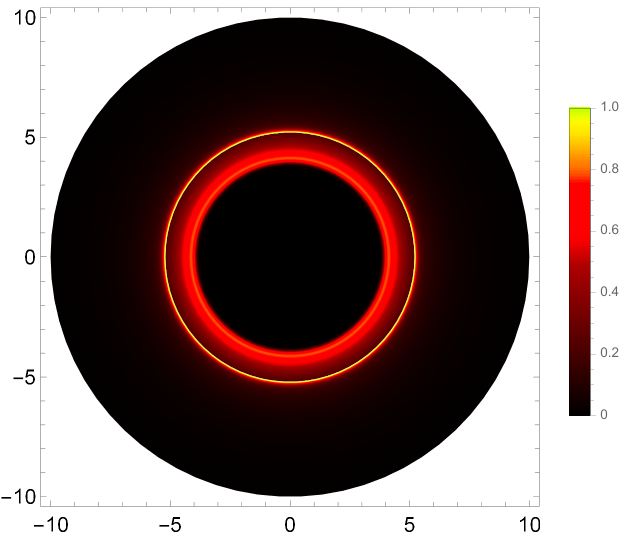}\qquad
    \includegraphics[scale=0.5]{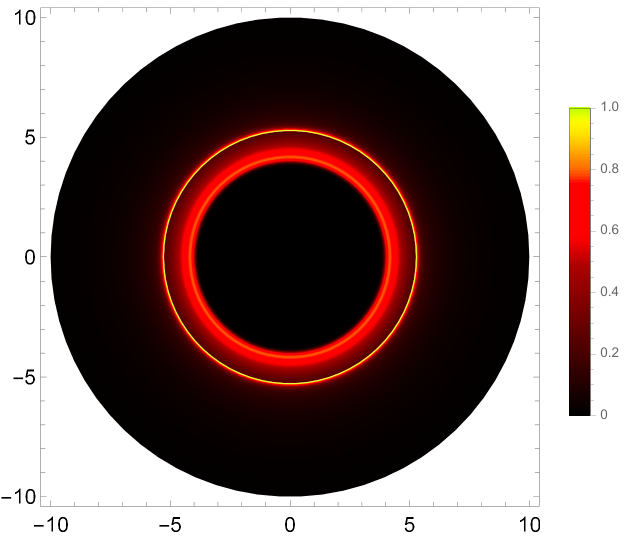}\\
    \includegraphics[scale=0.5]{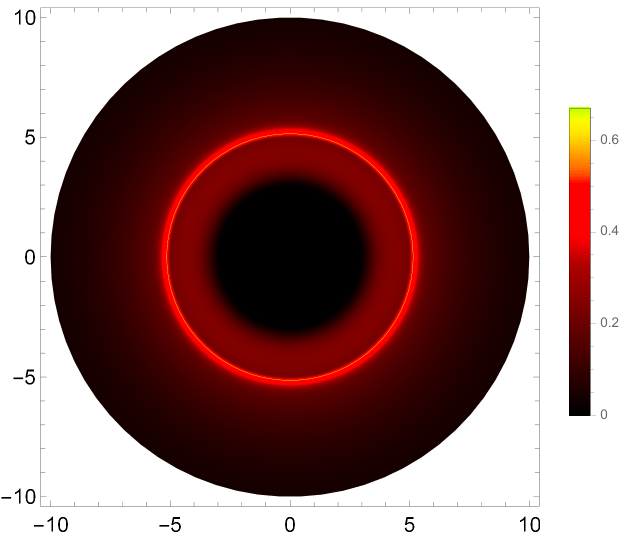}\qquad
    \includegraphics[scale=0.5]{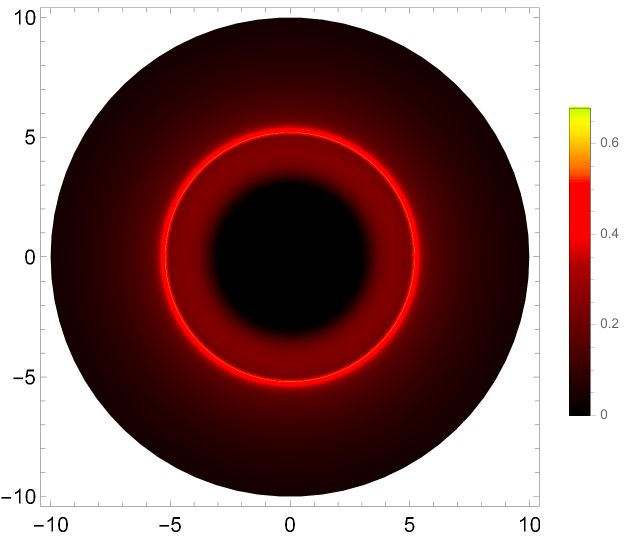}\qquad
    \includegraphics[scale=0.5]{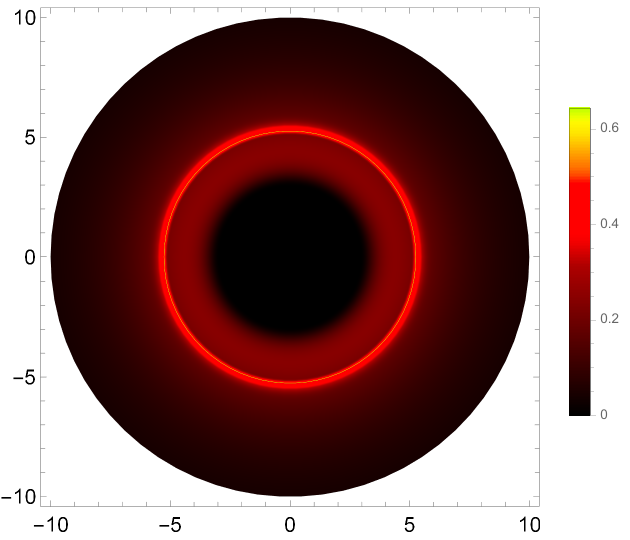}
    \caption{Shadow images for the ISCO disk model (top row), the LR disk model (middle row), and the EH disk model (bottom row), for $\alpha=-0.01$ (left column), $\alpha=0$ (middle column), and $\alpha=+0.01$ (right column).}
    \label{fig:shadows}
\end{figure*}

\begin{figure*}
\includegraphics[scale=0.7]{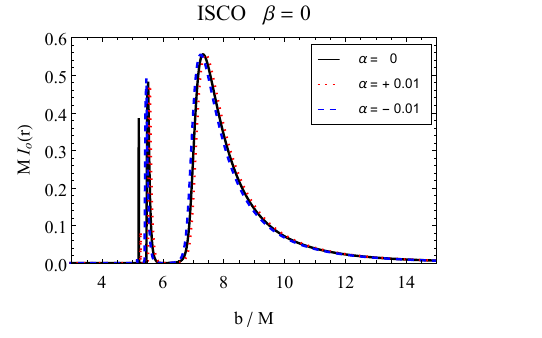}
\includegraphics[scale=0.7]{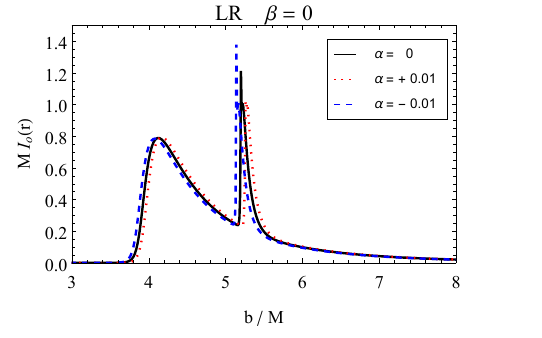}
\includegraphics[scale=0.7]{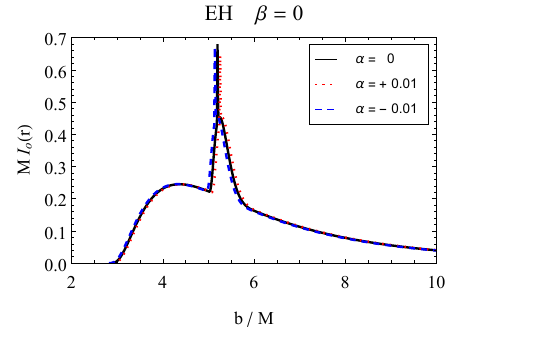}
\caption{Observed intensity $I(r)$ as a function of the radius $r$ for a varying $\alpha$ with constant $\beta=0$, and for the ISCO model (left panel), LR model (middle panel), and the EH model (right panel).}
\label{fig:intensities}
\end{figure*}

To analyse the effects of the parameters $\alpha$ and $\beta$ in the observational properties of the background spacetime given in Eq.\eqref{eq:metric} in the regime where the spherically symmetric approximation is valid, we have run the ray-tracing code for several combinations of setups. In particular, we have considered, $\alpha=\{-0.01; 0; 0.01\}$, and $\beta=\{-10^{-5}; 0; 10^{-5}\}$, for the three accretion disk luminosity profiles given in Table \ref{tab:disk_parameters}. The accretion disk is placed at the equatorial plane, i.e., at $\theta=\pi/2$, while the observer is placed at the vertical axis $\theta=0$ and at a radius of $r=600M$. Note that this approximation is valid only locally due to the alterations in the metric components caused by the parameter $\beta$. These alterations cause the spacetime to lose its asymptotic flatness, and thus the distance from the central object to the observer can not be made arbitrarily large. The value $r=600M$ was chosen in such a way that the effects of losing the asymptotic flatness are still negligible, while maintaining a distance large enough for the light rays to be approximately parallel when they reach the observer. In the equatorial plane validity region for $\beta<0$ is wider than for $\beta>0$, and an increase in $\beta$ leads to a decrease in the size of this region. We also chose relatively small values of $\beta$ with the goal of having a wider validity region allowing for the comparison of the results among different cases. Our results indicate that the effects of the parameter $\beta$ are negligible on the observed shadow image and luminosity profiles. Thus, in this section, we restrict our analysis to the effects of the parameter $\alpha$. For a complete set of results including the effects of the parameter $\beta$, we refer the reader to Appendix \ref{app:images}. The shadow images for the three disk models are presented in Figure \ref{fig:shadows}, whereas the observed intensity profiles for the three disk models are presented in Figure \ref{fig:intensities}. As for photon rings, their size and location can be quite different depending on the features of the parameters. In addition, the corresponding luminosity is also strongly related to the background geometry and the properties of the disk.

The first important feature to note is that the observed intensity profiles feature multiple peaks of intensity, in contrast to the emitted intensity profiles which feature a single peak of intensity, see Figure  \ref{fig:disks}. These multiple peaks of intensity at the observer are caused by photons that reach the observer after circling around the central object with a different number of half-orbits. In this peak structure, we can identify three main components: 

\begin{enumerate}
    \item the direct component: corresponding to the wider and dominant component, it is composed of the photons emitted directly from the accretion disk at $\theta=\pi/2$, i.e., the equatorial plane, to the observer at $\theta=0$. These photons have travelled a total of a quarter orbit around the central object, $\Delta\theta = \theta_e - \theta_o = \pi/2$, and are defined as the primary image;
    
    \item the lensed component: corresponding to a narrower component placed either inside (for the ISCO model) or outside (for the LR and EH models) the inner edge of the direct component, is composed of the photons that have been emitted by the accretion disk in the direction opposite to the observer, and then have been lensed around the central object once before reaching the observer. These photons have travelled an extra half-orbit than the photons of the direct component, $\Delta\theta = 3\pi/2$, and are defined as the secondary image;
    
    \item the photon ring component: corresponding to the narrowest of the three components, it is composed of the photons that have been emitted by the accretion disk and have orbited at least once around the central object, close to the LR, before reaching the observer \footnote{We note that, due to the limited resolution of our images and the fact that the photon ring component is the thinnest of the three observable components, the photon ring component should be used solely to determine the position of the critical curve in the observed image, and not to trace conclusions about the observed intensity.}. This component presents an infinite sub-structure of peaks corresponding to the photons that have orbited a total of $\Delta\theta=5\pi/2+n\pi$, for $n\geq 0$ an integer, which leads to additional brightness once again. This infinite sub-structure of peaks converges to critical curve $b=b_c$ which becomes exponentially closer to each other with an increase in the number of orbits. 
\end{enumerate}

For the ISCO model, these three components are all distinguishable in the observed emission profiles, see e.g., the top row in Figure \ref{fig:shadows} and the left panel in Figure \ref{fig:intensities}. However, for the LR model, the direct and lensed components appear superimposed, and for the EH model, the photon ring component is also superimposed with the direct emission.

In all three models considered for the accretion disk, one verifies that the parameter $\alpha$ affects the results more prominently than the parameter $\beta$. Indeed, since a variation in $\alpha$ induces a non-negligible translation of the radii of the ISCO and the LR of the background spacetime, one observes that the size of the shadow is strongly affected by this parameter, with the shadow radius increasing for $\alpha>0$ and decreasing for $\alpha<0$. Furthermore, one also verifies that the maximum amplitude of the observed intensity profiles for the direct emission is anti-correlated to the value of $\alpha$. This behaviour is expected as, for negative values of both quadrupole parameters, the radius of the ISCO is smaller than for the positive values \citep{2022Univ....8..195F}; thus, the radiation is being emitted closer to the central object. Indeed, for all of the accretion disk models, one verifies that the rate of change of the intensity profile as a function of the radial coordinate is more abrupt for larger values of $\alpha$, which is also expected. As a consequence, the lensed and photon ring components of the intensity profiles for negative values of alpha are wider, and thus more visible in the shadow images. On the other hand, the parameter $\beta$ presents a minor effect on the observational properties of this setting, inducing smaller changes in the observed intensity profiles in comparison with the effect of the parameter $\alpha$, see Appendix \ref{app:images}. Our results show that the observed intensity profiles are slightly closer to the central object for negative values of $\beta$, while positive values of beta tend to bring them farther. The effects are particularly clear for the ISCO model, for which the direct and lensed components of the observed intensity profile are separated, but it is also visible (although less contrasting) in the LR and EH models, for which the direct and lensed components of the observed intensity profile are superimposed. 

\begin{figure*}
    \centering
    \includegraphics[scale=0.89]{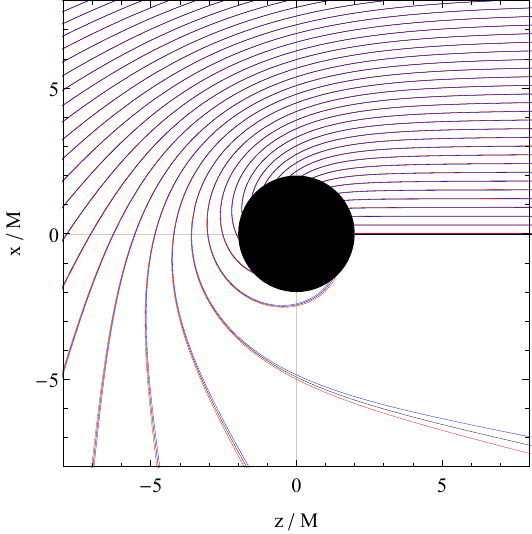}\qquad\qquad
    \includegraphics[scale=0.89]{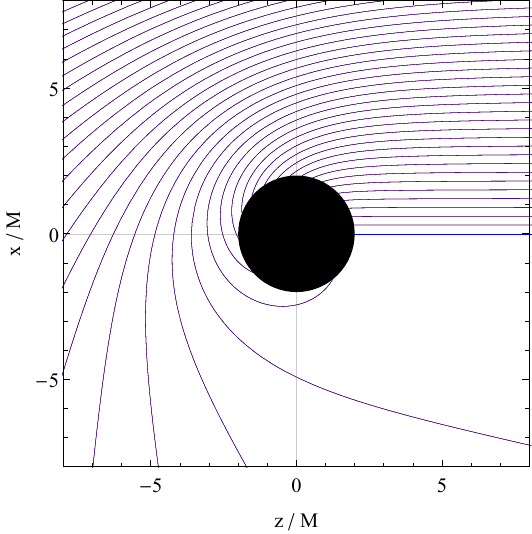}
    \caption{ Photon trajectories that reach a distant observer in the far right $z=100M$. The black disk at $x=1$ represents the event horizon for $\alpha=0$ or the naked singularity for $\alpha\neq 0$. Photon trajectories are colored according to the different parameters of the metric. Left panel: $\beta=0$ and prolate object $\alpha=-0.01$ (blue lines), Schwarzschild solution $\alpha=0$ (black lines) and oblate object $\alpha=0.01$ (red lines). Right panel: $\alpha=0$ and $\beta=-10^{-6}$ (blue lines), Schwarzschild solution $\beta=0$ (black lines), and $\beta=10^{-6}$ (red lines). Note that each set of three geodesics is emitted with the same impact parameter at $z=100M$ but, due to the fact that the spacetime is not asymptotically flat, they reach the vicinity of the central object visible on the plot with slight deviations from the asymptotically flat $\alpha=0$ or $\beta=0$ cases.}

    \label{fig:geodesics}
\end{figure*}

Figure \ref{fig:geodesics} shows how the null geodesic congruences are affected by parameters $\alpha$ and $\beta$. Although all calculations and plots extend up to an observer distance of $600M$, in this Figure we only display results out to $100M$ for convenience. The left panel shows the effect of $\alpha$ for a wide range of impact parameters. Most trajectories cross the equatorial plane once, but photons that propagate close to the unstable photon orbit on the image plane wrap around the central object and cross the plane twice, three times or more, with photons appearing exactly on the critical curve describing trajectories that asymptotically approach the unstable bound orbits composing the photon shell. 
The size of the shadow is defined by the range of impact parameters corresponding to the trajectories that do not cross the equatorial plane before intersecting the black disk.
Thus, one can see that the size of the shadow produced by an oblate compact object is larger than the shadow produced by the Schwarzschild blackhole, and the reverse is true for a prolate compact object. In the right panel, the impact of the parameter $\beta$ is shown. It is evident that the effect of the parameter $\beta$ in this setting and in the equatorial plane is much smaller in comparison with the effects of the parameter $\alpha$ as anticipated by the analysis of the intensity profiles and shadow images.

\section{Dark matter model} \label{darkmatter}

The Standard Model of Cosmology characterizes the Universe as geometrically flat, with dark energy, dark matter, and ordinary matter comprising approximately $68\%$, $27\%$, and $5\%$ of its total mass-energy density, respectively. Dark matter most likely consists of unknown elementary particles beyond the standard model of particle physics that originated in the early Universe, resulting in negligible effects of its velocity dispersion on structure formation (referred to as cold dark matter). Additionally, unlike baryonic matter, dark matter is not expected to form accretion disks due to its low interaction cross-section. Furthermore, due to the absence of the coupling with the electromagnetic field, light rays may travel through dark matter distributions virtually unaffected by any non-gravitational means\footnote{This is true, independently of the distribution and equation of state of the dark matter model.}. To date, there has been no direct detection of dark matter. The evidence supporting its existence is primarily based on cumulative measurements, such as galactic rotation curves \citep{1980ApJ...238..471R}, and the cosmic microwave background \citep{2014A&A...571A..16P}. In this study, we focus only on its fundamental property, more specifically, its manifestation through gravitational effects \citep{2019PhLB..795....1K}, through the definition of an additional effective mass function as

\begin{equation}
     m(r) =
    \begin{cases}
      M & \text{$r<r_s$} \\
      M+ \mathcal{S}(r)\Delta M & \text{$r_s\leq r\leq r_s+\Delta r$}\\
      M+ \Delta M & \text{$r_s+\Delta r<r$}
    \end{cases}       
\end{equation}
where $r_s=2M$ is the Schwarzschild radius, $\Delta M$ is the mass of the dark matter halo, $\Delta r$ is the thickness of the dark matter halo, and $\mathcal S$ is a radial function defined in such a way as to preserve the continuity of the mass function and its first derivative with respect to $r$. This function takes the form

\begin{align}
\mathcal{S}(r)= \left(3-2\frac{r-r_s}{\Delta r}\right)\left(\frac{r-r_s}{\Delta r}\right)^2.
\end{align}
In this definition, $\Delta M>0$ corresponds to the positive dark matter mass-energy density and $\Delta M<0$ to the negative one. In this work, we consider solely the positive case and replace $M$ by $m(r)$ in the line element in Eq. \eqref{eq:metric}. 

\begin{figure*} 
    \centering
    \begin{tabular}{cc}
    \includegraphics[width=0.4\hsize]{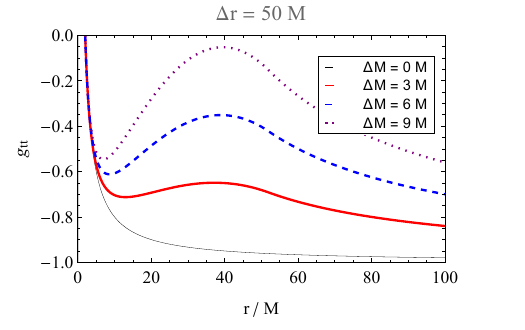}
    \includegraphics[width=0.4\hsize]{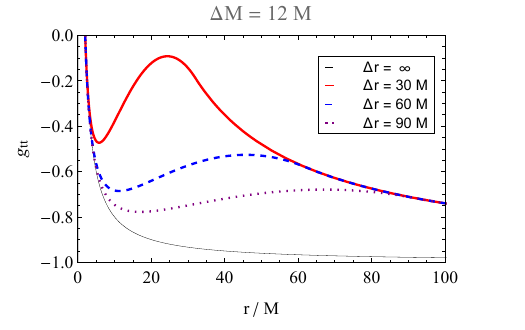}\\
    \includegraphics[width=0.4\hsize]{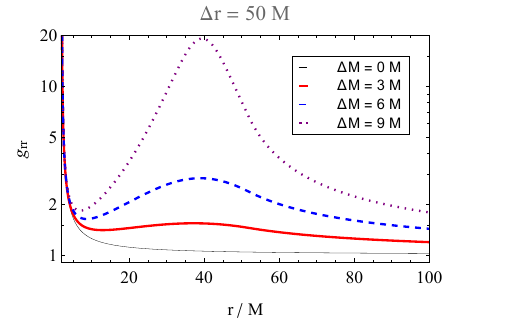}
    \includegraphics[width=0.4\hsize]{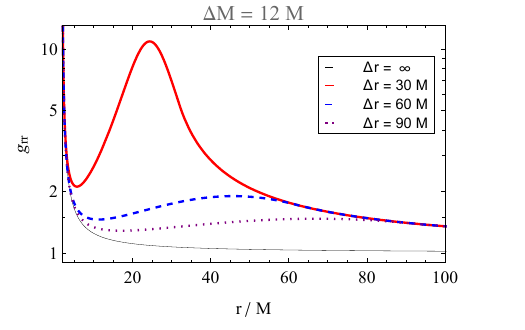}
    \end{tabular}
    \caption{Metric components $g_{tt}$ (first row) and $g_{rr}$ (second row) for the dark matter model with $\Delta r=50M$ with varying $\Delta M$ (left plot) and for $\Delta M=12M$ with varying $\Delta r$ (right plot), with $\alpha=\beta=0$.}  
    \label{DMmetric1}
\end{figure*}

Figure \ref{DMmetric1} shows the behaviour of the $g_{tt}$ and $g_{rr}$ components for the dark matter model for different combinations of $\Delta M$ and $\Delta r$. To see the effect of the dark matter distribution on the spherically symmetric black hole case, we consider the combination $\alpha=\beta=0$. Compared to the vanishing dark matter case, we see that the metric coefficients have two additional local extrema, a local minimum close to the inner edge of the disk corresponding to the inner edge of the dark matter halo, and a local maximum at the outer edge of the halo. Furthermore, the magnitude of these extrema is anti-correlated to the thickness of the halo, with large values of $\Delta r$ inducing only slight modifications to the spacetime geometry close to the central object. Furthermore, previous publications have shown that, under certain conditions, the spacetime presents an additional pair of unstable photon orbits with opposite stability characters \citep{2019PhLB..795....1K}. Note that if $r_o \leq r_{LR} < r_s$, where $r_o$ is the radius at which the observer stands, the observer can only see the purely compact object effect on the shadow without the contribution of dark matter. Also, if $r_s + \Delta r < r_{LR} < r_o$, the dark matter halo is constrained in the vicinity of the central object but it is assumed not to fall into the event horizon, which is a rather nonphysical assumption. We thus choose to discard the two cases above and restrict our analysis to the combination $r_s < r_{LR} < r_s + \Delta r$. 

\begin{figure*}
\centering
\includegraphics[scale=0.9]{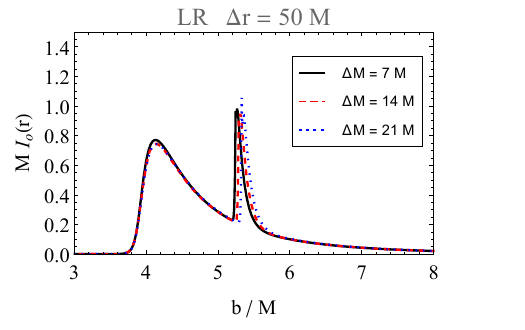}
    \includegraphics[scale=0.9]{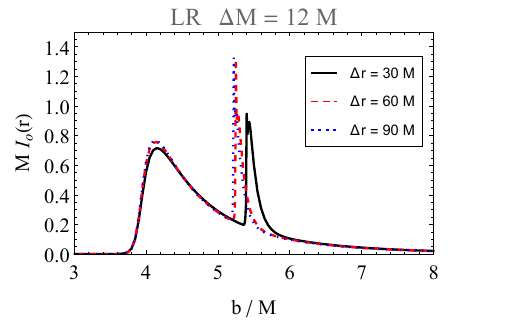}\\
    \includegraphics[scale=0.9]{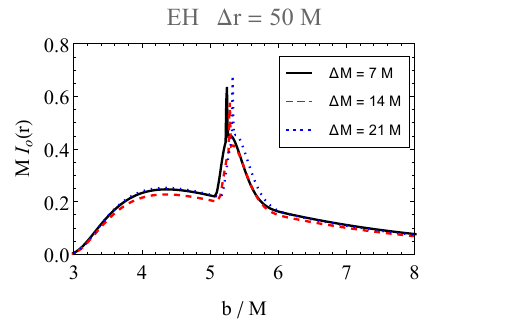}
    \includegraphics[scale=0.9]{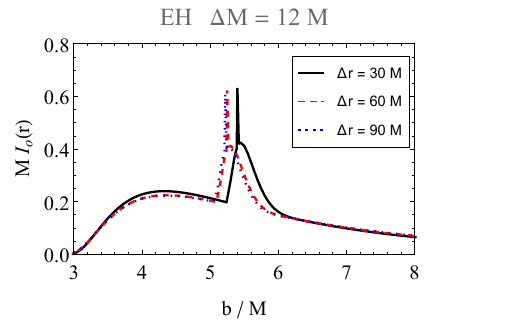}
    \caption{Intensity $I(r)$ as a function of the radius $r$ for the LR disk model (top row) and for the EH disk model (bottom row), with $\alpha=\beta=0$, for $\Delta r= 50M$ with $\Delta M=\{7 M, 14M, 21M\}$ (left column), and $\Delta M= 12M$ with $\Delta M=\{30 M, 60M, 90M\}$ (right column).}
    \label{fig:intensity_DM1}
\end{figure*}

\begin{figure*}
\centering
\includegraphics[scale=0.5]
{plot_a0_b0_LR.pdf}
\includegraphics[scale=0.5]{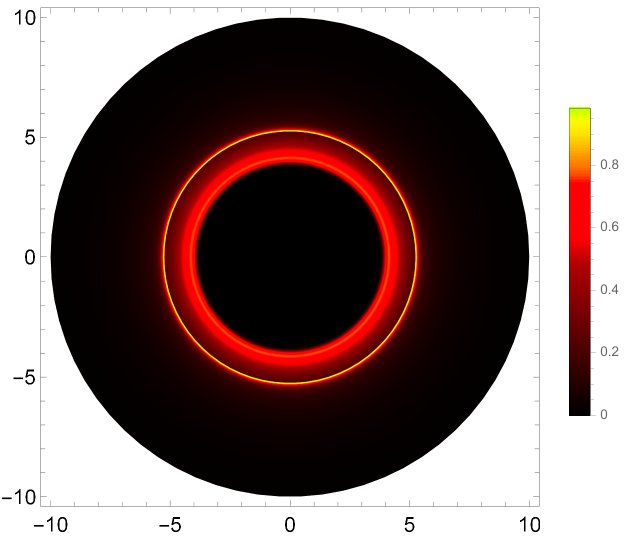}
\includegraphics[scale=0.5]{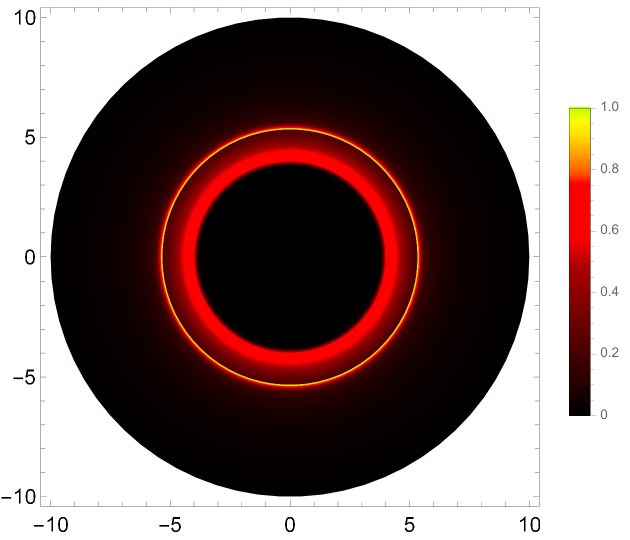}\\
\includegraphics[scale=0.5]{plot_a0_b0_EH.pdf}
\includegraphics[scale=0.5]
{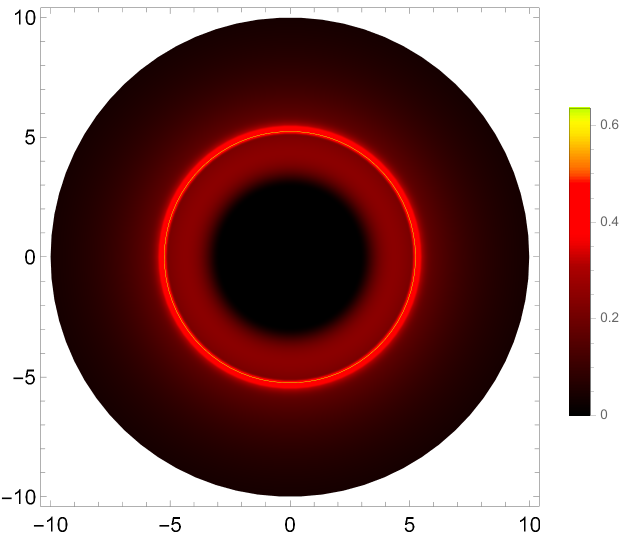}
\includegraphics[scale=0.5]{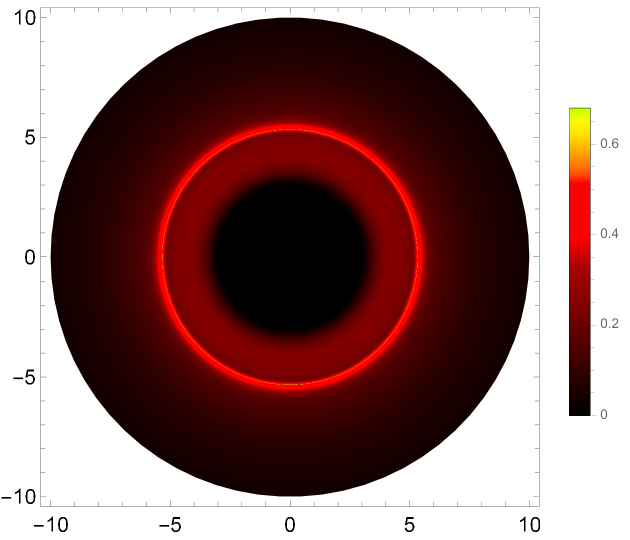}\\
\caption{Shadow images for the LR disk model (top row) and EH model (bottom row) with $\alpha=\beta=0$. The first column depicts the absence of dark matter, while the second and third columns show $\Delta r= 50M$ and variations in $\Delta M={7M, 21M}$, respectively.}
\label{fig:shadows_DM_LREH}
\end{figure*}

The methods outlined in the previous section for the ray-tracing and production of the observed intensity profiles and shadow images were also applied to the model introduced in this section. However, since the radius of the ISCO is highly sensitive to the mass distribution of the halo and it is a discontinuous function of $\Delta M$ for configurations in which the mass of the halo is much larger than the mass of the central object, we have decided to restrict our analysis to the LR and EH models for the accretion disk. The observed intensity profiles $I(r)$ for the particular case $\alpha=\beta=0$ for the LR and EH models are given in Figure \ref{fig:intensity_DM1}. We chose $\alpha=\beta=0$ to analyse separately the effect of the dark matter model on the intensity profiles. Our results indicate that both the LR and the EH models are only slightly sensitive to both $\Delta M$ and $\Delta r$, with variations in the smaller values of $\Delta r$ and larger values of $\Delta M$ being the most prominent. Figure \ref{fig:shadows_DM_LREH} shows the shadow images for the LR and EH disk models for the same particular case $\alpha=\beta=0$. This figure compares the shadow images in the absence of dark matter with two mass distributions of the same length-scale with different masses. One can observe a slight increase in the size of the shadow for both disk models with an increase in the mass of the halo. 

\begin{figure*}
\centering
 \begin{tabular}{cc}
    \includegraphics[width=0.5\hsize]{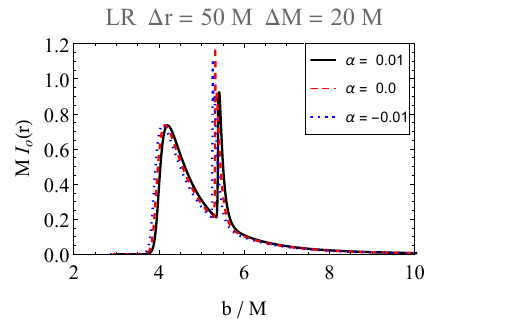}&
    \includegraphics[width=0.5\hsize]{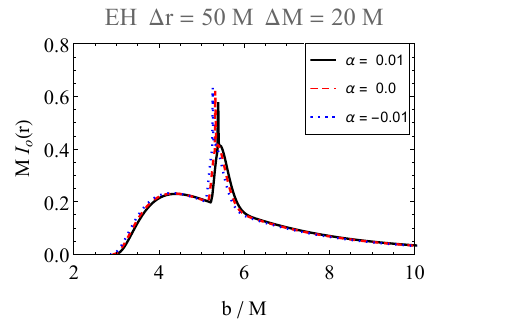}
     \end{tabular}
    \caption{Intensity $I(r)$ as a function of the radius $r$ for the LR disk model (left plot) and for the EH disk model (right plot), with $\Delta M=20M$, $\Delta r=50 M$ for $\alpha=\{-0.01, 0, 0.01\}$ and $\beta=0$.}
    \label{fig:intensity_DM2}
\end{figure*}

\begin{figure*}
    \includegraphics[width=0.5\hsize]{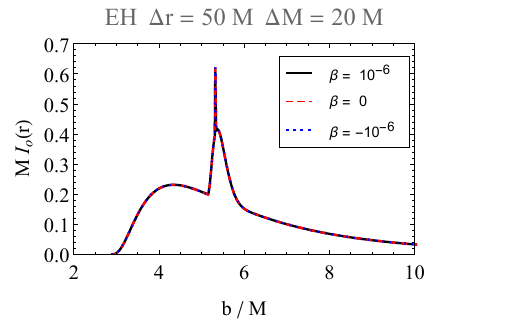}
    \includegraphics[width=0.5\hsize]{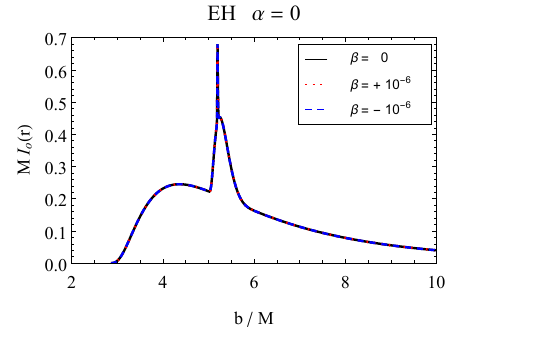}
    \caption{Comparing the intensity $I(r)$ for the ER disk model with $\alpha=0$ and varying $\beta$, both in the presence and absence of dark matter. }
    \label{fig:intensity_DM3}
\end{figure*}

\begin{figure*}
\centering
\includegraphics[scale=0.5]{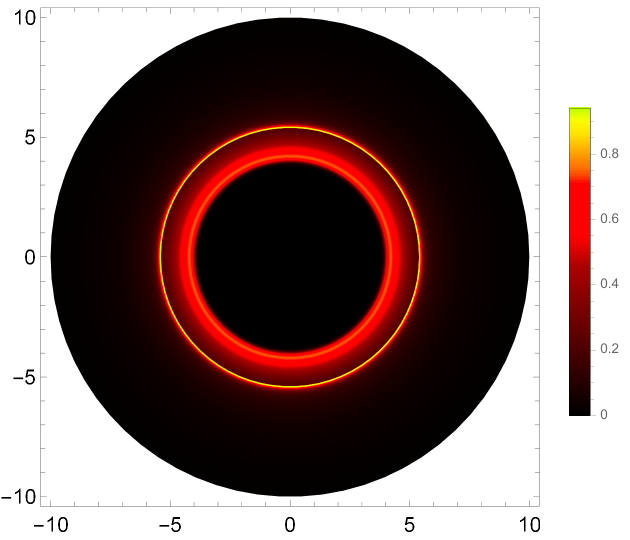}
\includegraphics[scale=0.5]{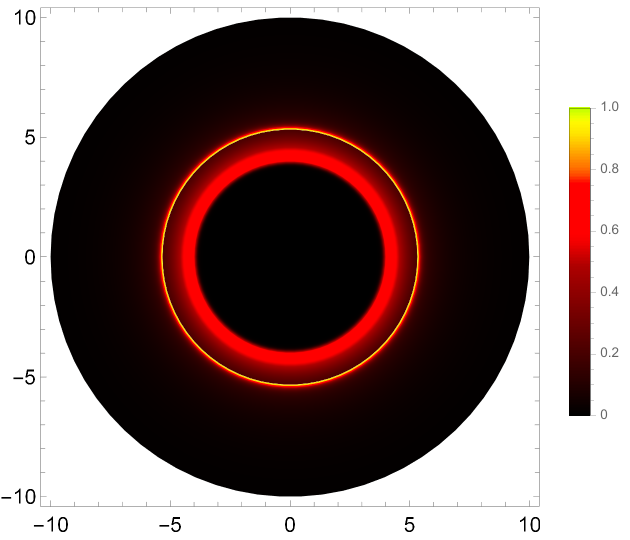}
\includegraphics[scale=0.5]{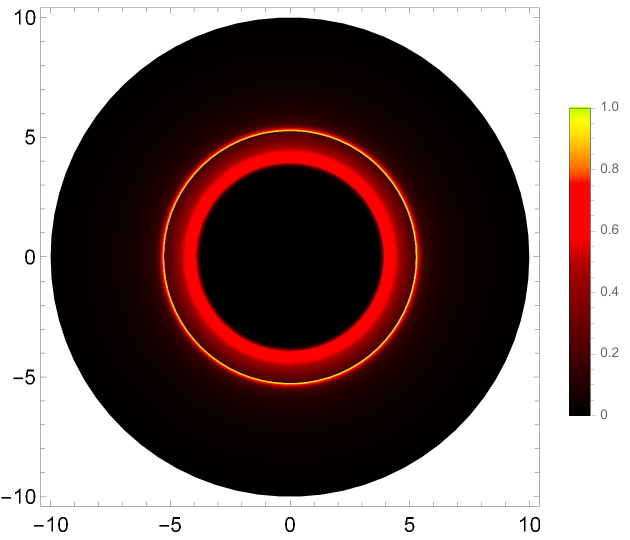}\\
\includegraphics[scale=0.5]{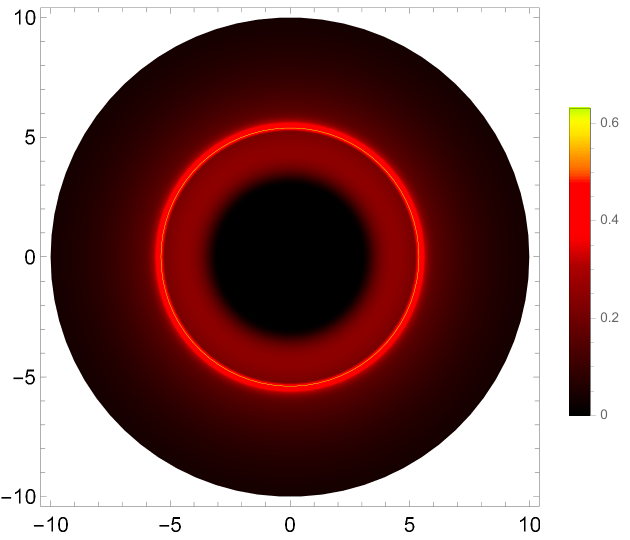}
\includegraphics[scale=0.5]{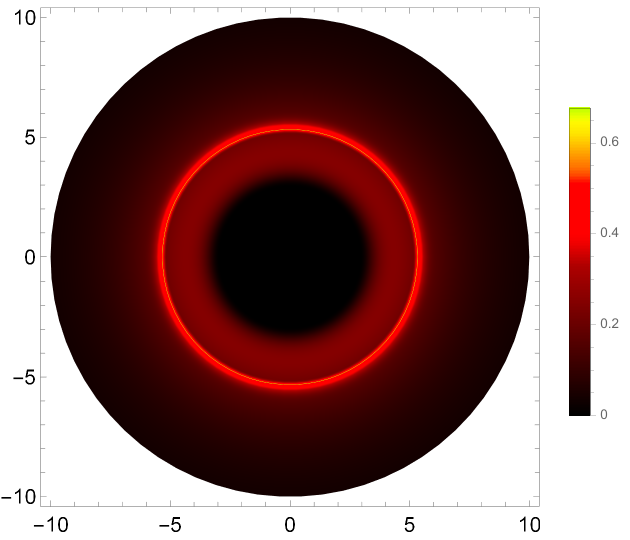}
\includegraphics[scale=0.5]{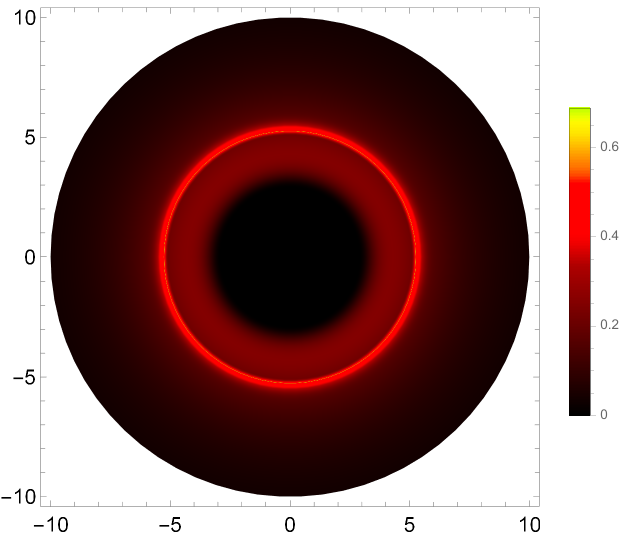}
\caption{Shadow images for the LR disk model (top row) and the EH disk model (bottom row) with $\Delta M=20M$ and $\Delta r=50M$, for $\alpha=0.01$ (left column), $\alpha=0$ (middle column), and $\alpha=-0.01$ (right column).}
\label{fig:shadows_DM_LR_alpha}
\end{figure*}

Turning now to the analysis of how the parameters $\alpha$ and $\beta$ influence the results, Figure \ref{fig:intensity_DM2} shows the observed intensity profiles for varying $\alpha$ with $\beta=0$ for a given set of parameters of the dark matter model. Similarly to what was observed in the absence of the DM halo in Figure \ref{fig:shadows} and Figure \ref{fig:intensities}, variations in $\alpha$ induce radial translations of the observed intensity profiles. This is due to the fact that the accretion disk is closer to the compact object for smaller values of $\alpha$. Furthermore, upon comparing the LR panel in Figure \ref{fig:intensities} and Figure \ref{fig:intensity_DM2} for the cases in which $\beta=0$, it is evident that for any constant value of $\alpha$, the radial position of the peak shifts to larger values. We see the same pattern in the EH model panels of Figure \ref{fig:intensity_DM2} and Figure \ref{fig:intensities}. Furthermore, in accordance with the results of the previous section, we observe that variations in the chosen values for the parameter $\beta$ induce barely noticeable changes in the observed intensity profiles. To clarify this point, in Figure \ref{fig:intensity_DM3} we show the observed intensity profiles for $\alpha=0$ for both the DM model and in the absence of DM with varying $\beta$. These results indicate that the presence of DM does not alter the sub-dominant character of the parameter $\beta$ in the observational properties of these models. In Figure \ref{fig:shadows_DM_LR_alpha}, we depicted the shadow images for varying $\alpha$ and vanishing $\beta$ for EH and LR models. The results are comparable with those of Figure \ref{fig:intensity_DM2}.

Summarising, our results indicate that considering DM model that we chose in this work, the presence of DM halos affects only slightly the observed intensity profiles and shadows of the models considered, thus being a sub-dominant effect in comparison with the effects of the dominant deformation parameter $\alpha$. Nevertheless, for a given model with a set value of the deformation parameter, the effects of the DM halo increase with the mass and the proximity of the halo from the central compact object.

\section{Summary and Conclusion} \label{summary}

In this work, we have analysed the shadow images and photon ring structure of a novel class of compact objects that are analytically tractable, with the goal of extending the previous research conducted in different spacetimes. This spacetime is a solution to the Einstein field equation and an extension of the familiar $\rm q$-metric, which is static and axisymmetric, and includes a distortion parameter $\beta$ and a deformation parameter $\alpha$, as briefly outlined in Section \ref{spacetime}. This analysis was done both in the absence and the presence of a dark matter halo, the latter represented by a modification in the radial mass function with two free parameters controlling the mass $\Delta M$ and the typical length-scale $\Delta r$ of the halo.

In our analysis, we have assumed the gas medium to be optically thin. The distinctive aspects of the observed images are primarily influenced by the location and characteristics of the emitting material close to the central object. In general, it can be inferred that the inclusion of a quadrupole in the metric significantly impacts the size and intensity of shadow images. Specifically, the oblateness of the central compact object, represented by the $\alpha$ parameter, plays a more pronounced role in influencing the characteristics of the shadow. This is due to the fact that modifications in $\alpha$ induce large variations in the radius of the ISCO and the unstable photon orbit, which control the interior edge of the accretion disk models. On the other hand, we have shown that the distortion parameter $\beta$ plays a negligible role in affecting the observational properties of these spacetimes. These results indicate that the analysis of the sizes of shadows observed experimentally could be used to constraint the oblateness of the central compact object, but in the absence of oblateness, one can distinguish the effect of distortion due to the external matter distribution.

Furthermore, we have analysed the effects of the presence of a dark matter halo surrounding the central object and the possibility of identifying this presence from the physical properties of emission from an optically thin disk region. Our results indicate that dark matter has the ability to either enhance or diminish the size of the observed shadow, although these effects are sub-dominant in comparison with the effects of oblateness, considering the chosen parameter's range. Note that this outcome is contingent upon the specific dark matter model being considered. In this work, we have considered a rather simple model for the dark-matter halo, however, one can consider more complicated models e.g., \citep{1990ApJ...356..359H,2003gr.qc.....3031K,PhysRevD.86.123015,2020PhRvD.101b4029X} for further investigation.

The work considered in this manuscript provides a starting step to the analysis of more realistic and astrophysically motivated models. Indeed, several assumptions taken in this work could be dropped. Larger variations of the deformation and distortion parameters, for which the spherically symmetric approximation ceases to be valid, could be considered. Furthermore, in a realistic astrophysical scenario, we expect compact objects to be rotating, in contrast with the static configurations considered here. Finally, our analysis assumed the observer to be face-on with respect to the equatorial plane. While this assumption is acceptable if one wants to perform a comparison with the experimental data from M87, a generalization for inclined observations could be beneficial to allow for a comparison with the Sgr A* data. We leave these extensions for future publications.

\begin{acknowledgements}
Sh.F. is funded by the University of
Waterloo and in part by the Government of Canada through the Department of Innovation, Science and Economic Development and by the Province of Ontario through the Ministry of Colleges and Universities at Perimeter Institute. Also, thanks the excellence cluster QuantumFrontiers funded by the Deutsche Forschungsgemeinschaft (DFG, German Research Foundation) under Germany’s Excellence Strategy – EXC-2123 QuantumFrontiers.
J.L.R. acknowledges the European Regional Development Fund and the programme Mobilitas Pluss for financial support through Project No.~MOBJD647, project No.~2021/43/P/ST2/02141 co-funded by the Polish National Science Centre and the European Union Framework Programme for Research and Innovation Horizon 2020 under the Marie Sklodowska-Curie grant agreement No. 94533, Fundação para a Ciência e Tecnologia through project number PTDC/FIS-AST/7002/2020, and Ministerio de Ciencia, Innovación y Universidades (Spain), through grant No. PID2022-138607NB-I00.

\end{acknowledgements}

\bibliographystyle{aa}
\bibliography{shadowpq}

\newpage

\begin{appendix} 

\section{Validity of the spherical symmetry approximation} \label{app:approx}

\begin{figure*}
    \centering
    \includegraphics[width=0.3\linewidth]{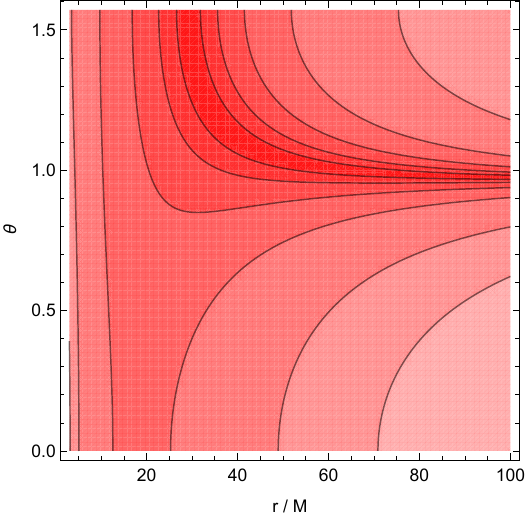}\quad
    \includegraphics[width=0.3\linewidth]{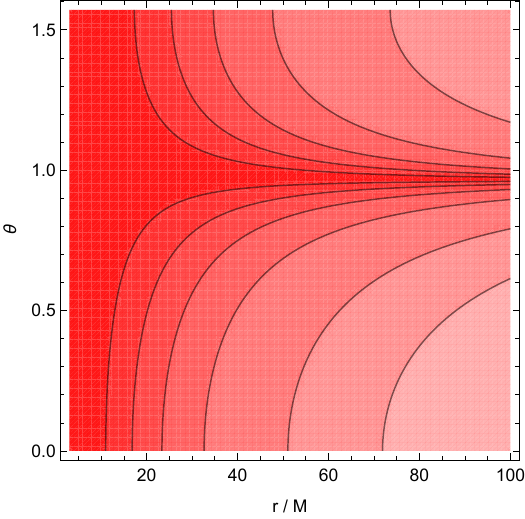}\quad
    \includegraphics[width=0.3\linewidth]{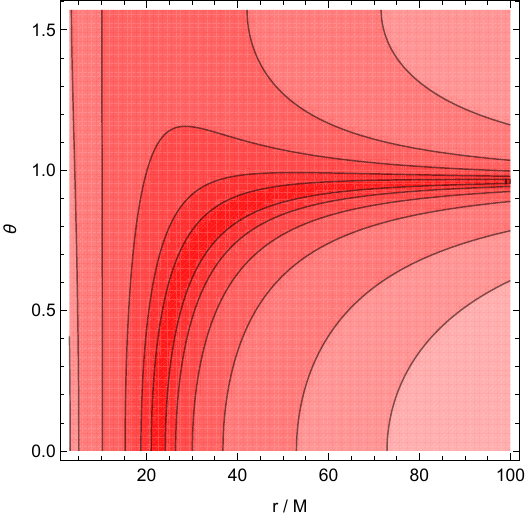}\\
    \includegraphics[width=0.3\linewidth]{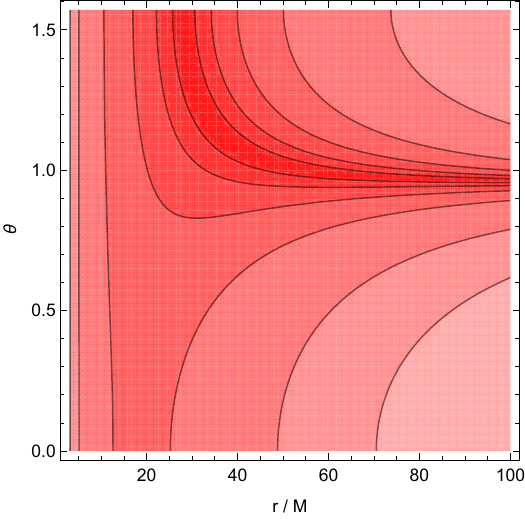}\quad
    \includegraphics[width=0.3\linewidth]{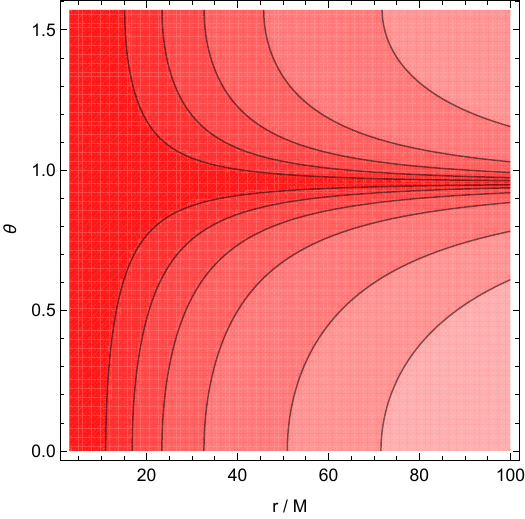}\quad
    \includegraphics[width=0.3\linewidth]{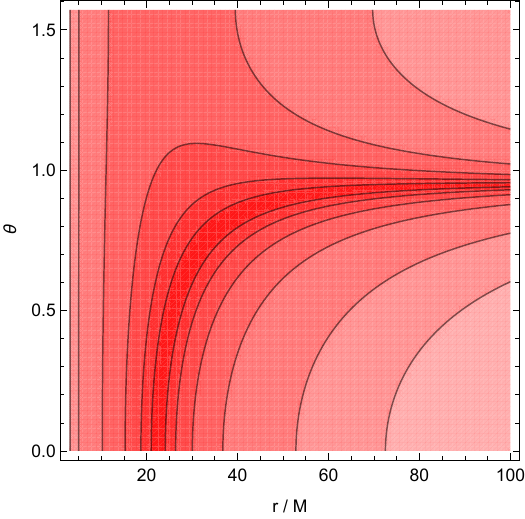}\\
    \includegraphics[width=0.3\linewidth]{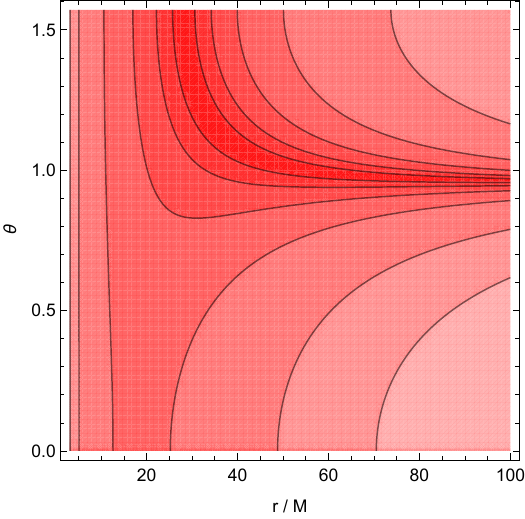}\quad
    \includegraphics[width=0.3\linewidth]{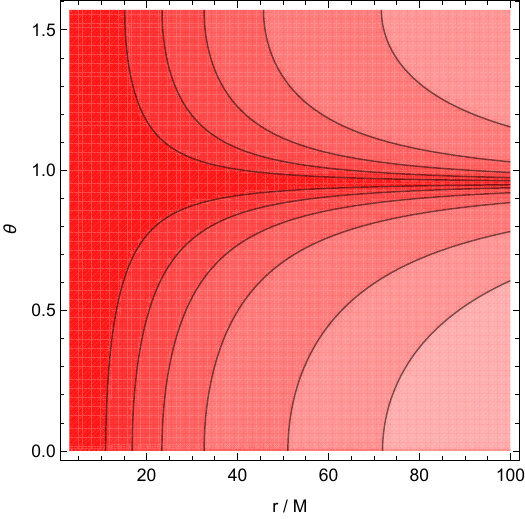}\quad
    \includegraphics[width=0.3\linewidth]{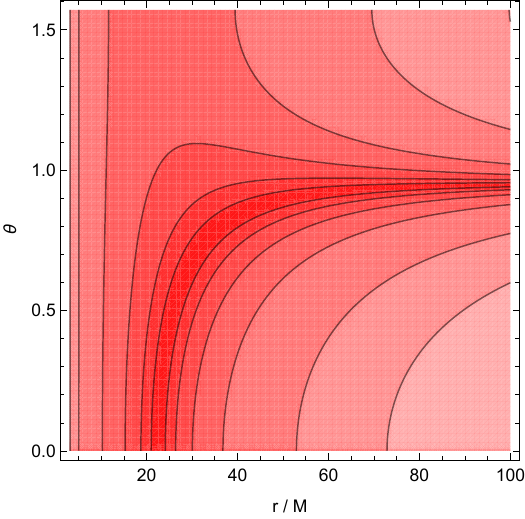}
    \caption{Relative errors $\nu_r$ (top row), $\nu_t$ (middle row), and $\nu_\phi$ (bottom row) as defined in Eq. \eqref{eq:errors} for $\alpha=\mp 0.01$ and $\beta=\pm 10^{-6}$ (left column), $\alpha=0$ and $\beta=\pm 10^{-6}$, and $\alpha=\pm0.01$ and $\beta=\pm10^{-6}$. The shades represent, from darker to lighter, the regions where the errors are below $0.02\%$, $0.05\%$, $0.1\%$, $0.2\%$, $0.5\%$, and $1\%$, with the largest errors being of the order of $2\%$.}
    \label{fig:errors}
\end{figure*}

In this section, we prove that the approximation taken in this work, i.e., that one can integrate the geodesic equation and obtain the trajectory of the light rays by taking the spacetime to be approximately spherically symmetric. For this purpose, and since the geodesic equation for the photons depends on the $g_{rr}$, $g_{tt}$, and $g_{\phi\phi}$ components of the metric, we calculate the relative errors between these metric components of the spherically symmetric Schwarzschild solution (defined by the parameters $\alpha=0$ and $\beta=0$) and those of the solutions with non-zero $\alpha$ and $\beta$. We define the relative error as:
\begin{equation}\label{eq:errors}
    \nu_i = \left|\frac{g_{ii}^S-g_{ii}}{g_{ii}^S}\right|,
\end{equation}
where $\nu$ is the relative error, the subscript $i$ can denote either $r$, $t$, or $\phi$, and $g_{\mu\nu}^S$ is the Schwazrschild spacetime metric. In the region of interest where photons are propagating between the observer and the central object and not falling into the event horizon, i.e., for a range $r\in \left[3M,100M\right]$ and $\theta=\left[0,\frac{\pi}{2}\right]$, the relative errors are plotted in Figure \ref{fig:errors}. These results show that, in the regions where the photons propagate, the relative errors are consistently smaller than $1\%$, thus validating the approximation used.

\section{Complete shadow images and intensity plots}\label{app:images}

\begin{figure*}
    \centering
    \includegraphics[scale=0.5]{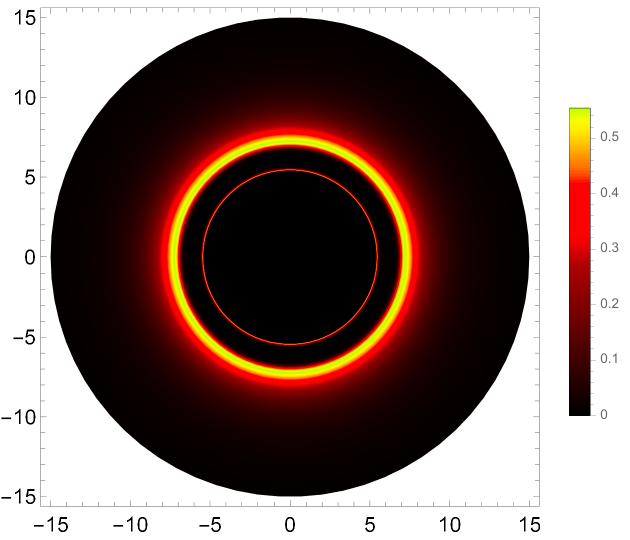}
    \includegraphics[scale=0.5]{plot_am3_b0_ISCO.pdf}
    \includegraphics[scale=0.5]{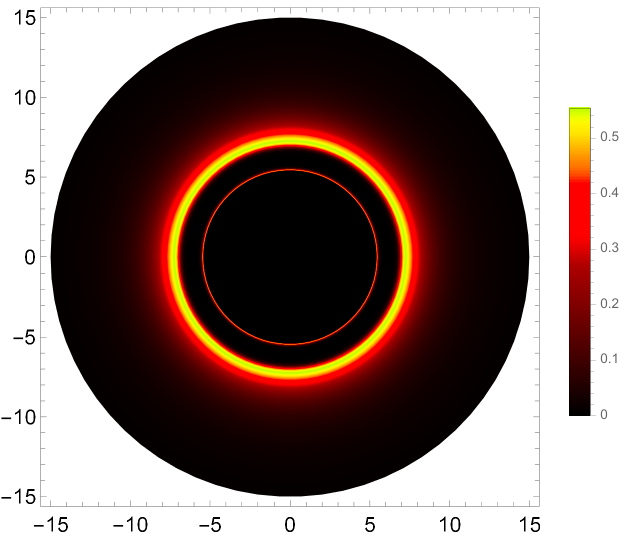}\\
    \includegraphics[scale=0.5]{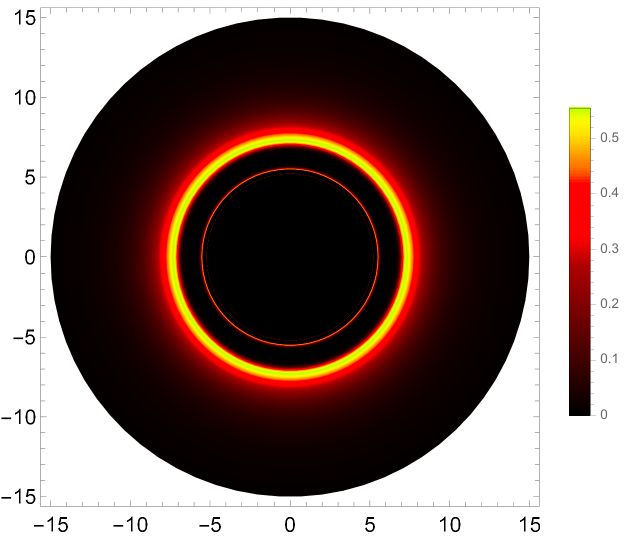}
    \includegraphics[scale=0.5]{plot_a0_b0_ISCO.pdf}
    \includegraphics[scale=0.5]{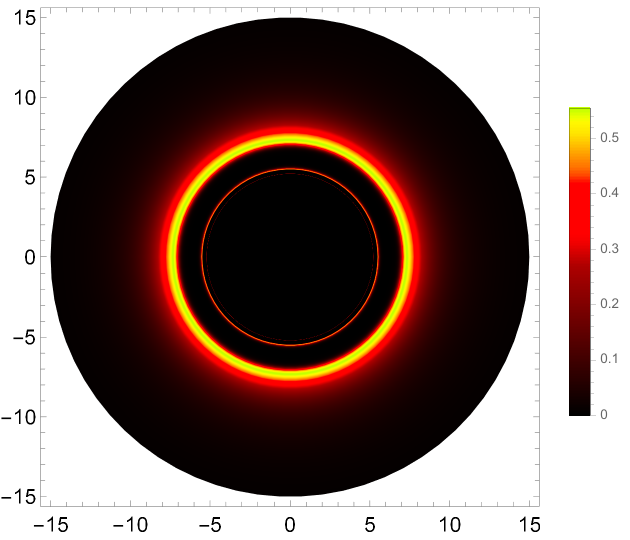}\\
    \includegraphics[scale=0.5]{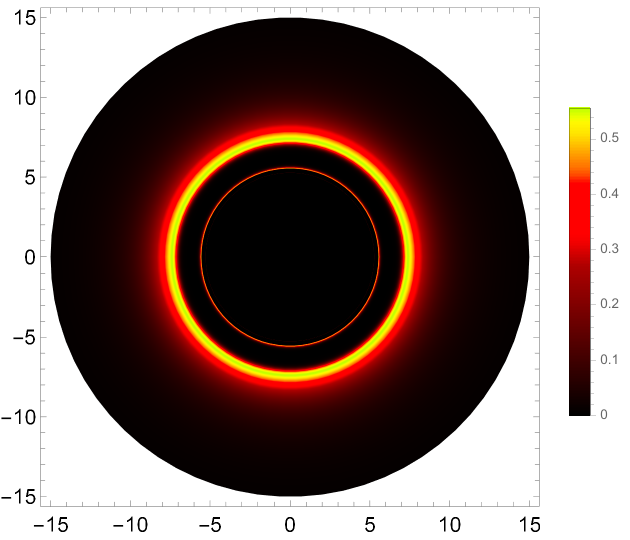}
    \includegraphics[scale=0.5]{plot_ap3_b0_ISCO.pdf}
    \includegraphics[scale=0.5]{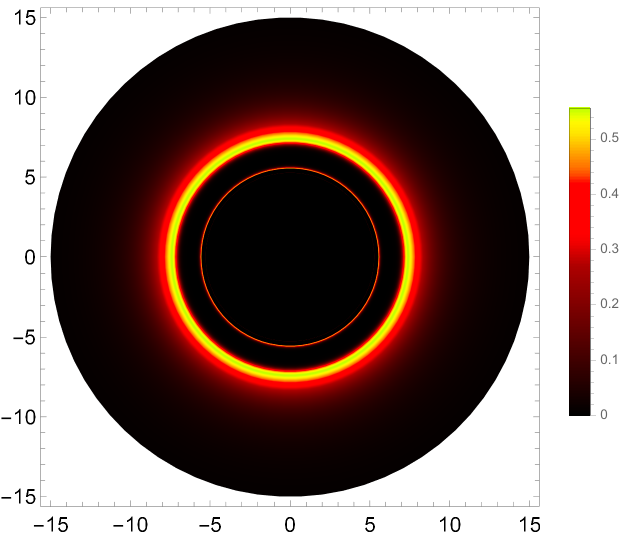}\\
    \caption{Shadow images for the ISCO disk model with different combinations of $\alpha$ and $\beta$, namely $\alpha=-0.01$ (top row), $\alpha=0$ (middle row), and $\alpha=+0.01$ (bottom row); and $\beta=-10^{-5}$ (left column), $\beta=0$ (middle column), and $\beta=10^{-5}$ (right) column). }
    \label{fig:shadows_ISCO_A}
\end{figure*}

\begin{figure*}
\includegraphics[scale=0.7]{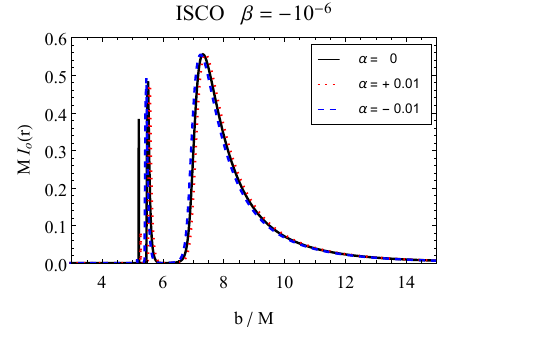}
\includegraphics[scale=0.7]{int_b0_ISCO.pdf}
\includegraphics[scale=0.7]{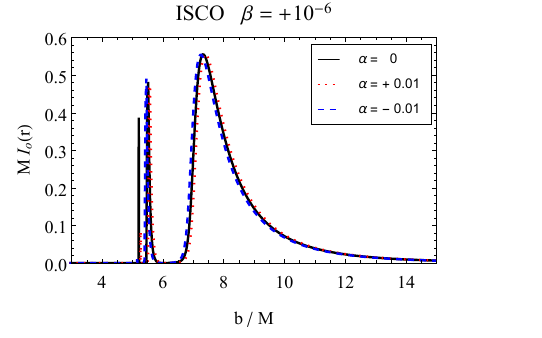}
\includegraphics[scale=0.7]{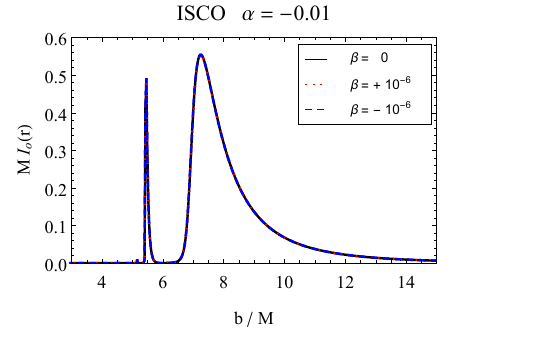}
\includegraphics[scale=0.7]{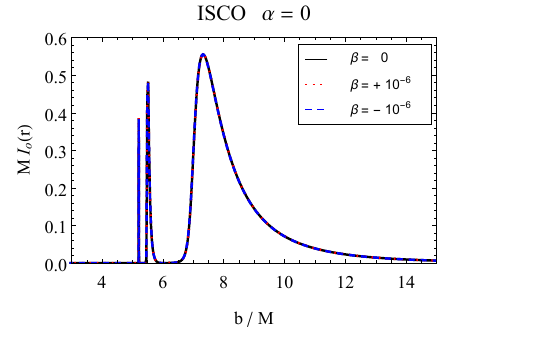}
\includegraphics[scale=0.7]{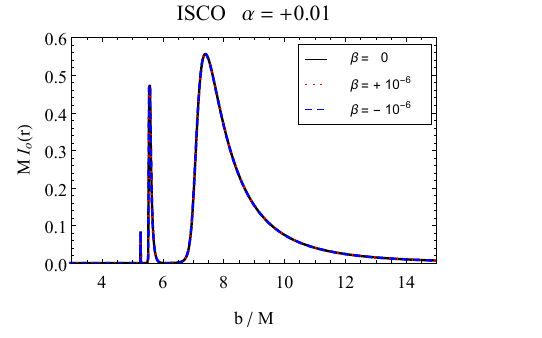}
\caption{Intensity $I(r)$ as a function of the radius $r$ for the ISCO disk model for a varying $\alpha$ with constant $\beta$ (top row) and for a varying $\beta$ with constant $\alpha$ (bottom row)}
\label{fig:intensity_ISCO_A}
\end{figure*}

\begin{figure*}
    \centering
    \includegraphics[scale=0.5]{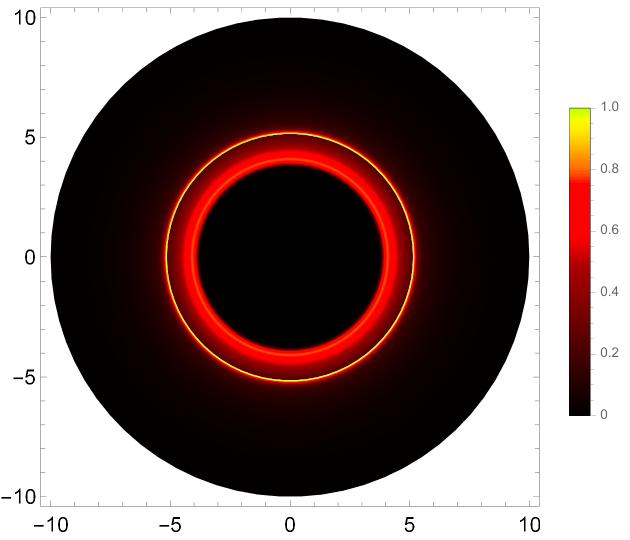}
    \includegraphics[scale=0.5]{plot_am3_b0_LR.pdf}
    \includegraphics[scale=0.5]{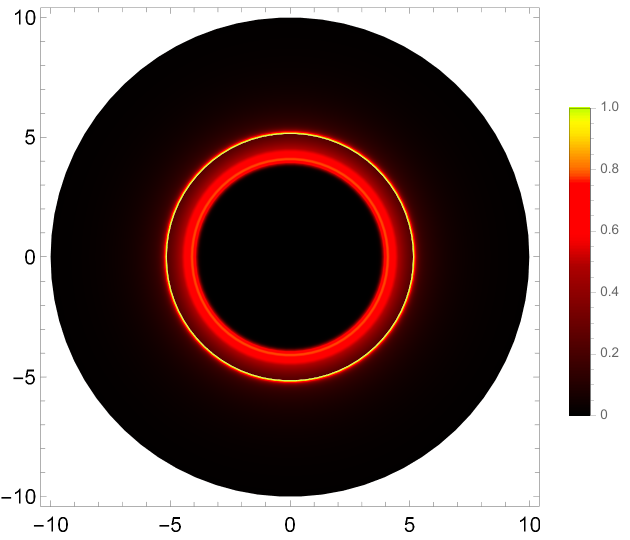}\\
    \includegraphics[scale=0.5]{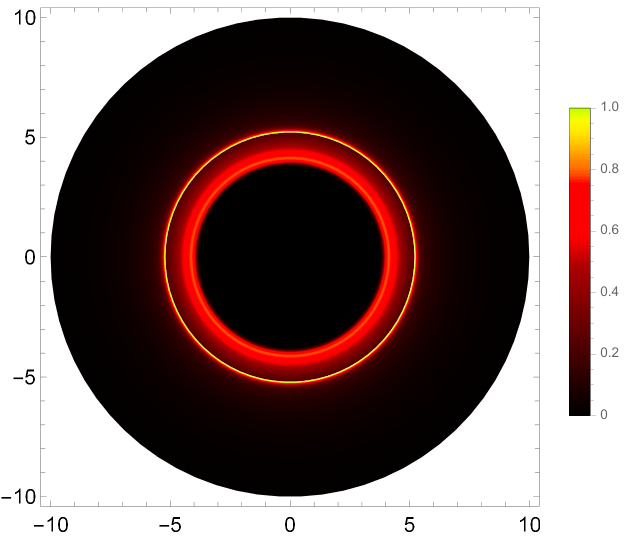}
    \includegraphics[scale=0.5]{plot_a0_b0_LR.pdf}
    \includegraphics[scale=0.5]{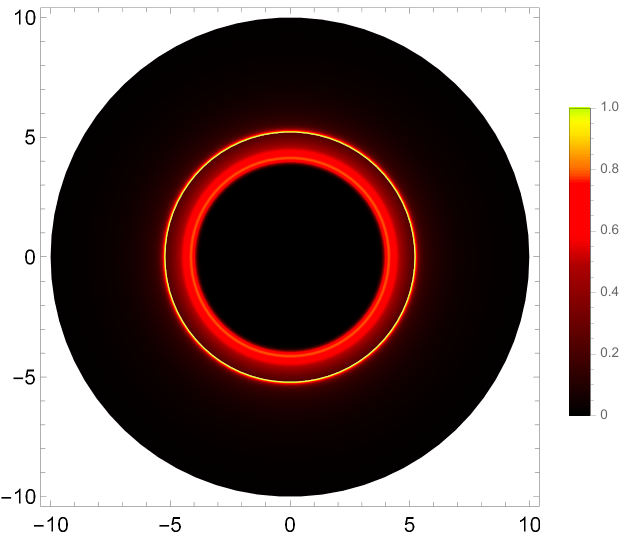}\\
    \includegraphics[scale=0.5]{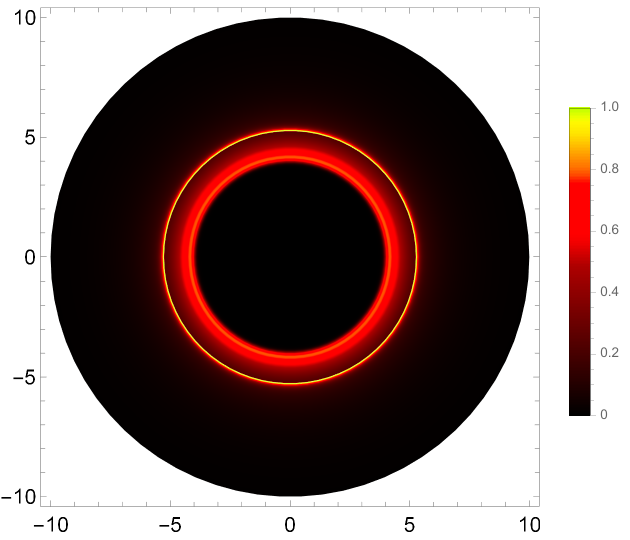}
    \includegraphics[scale=0.5]{plot_ap3_b0_LR.pdf}
    \includegraphics[scale=0.5]{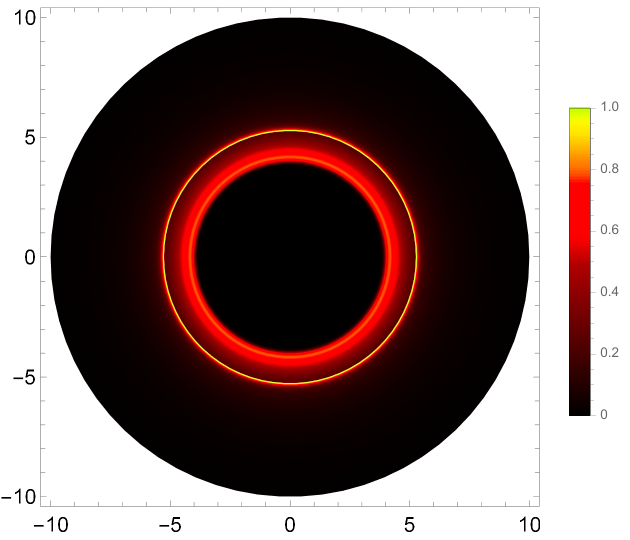}\\
    \caption{Shadow images for the LR disk model with different combinations of $\alpha$ and $\beta$, namely $\alpha=-0.01$ (top row), $\alpha=0$ (middle row), and $\alpha=+0.01$ (bottom row); and $\beta=-10^{-5}$ (left column), $\beta=0$ (middle column), and $\beta=10^{-5}$ (right) column). }
    \label{fig:shadows_LR_A}
\end{figure*}

\begin{figure*}
\includegraphics[scale=0.7]{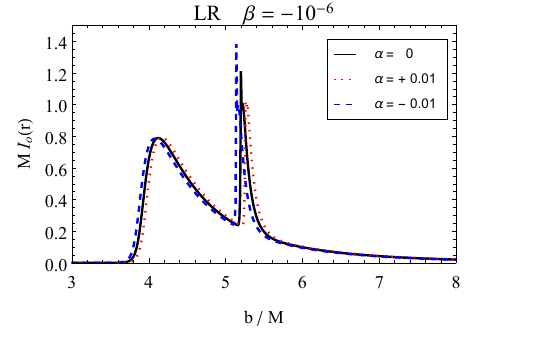}
\includegraphics[scale=0.7]{int_b0_LR.pdf}
\includegraphics[scale=0.7]{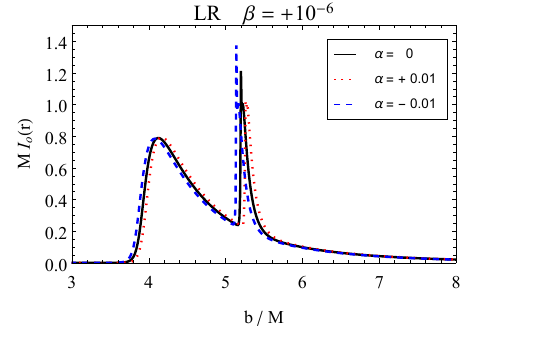}
\includegraphics[scale=0.7]{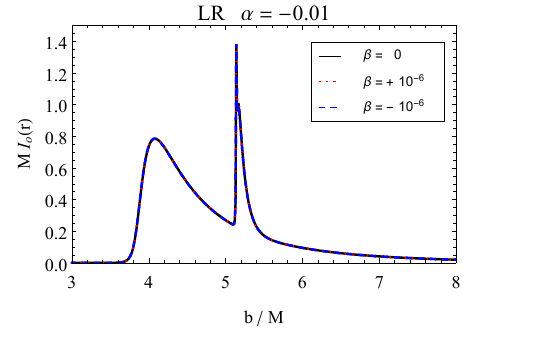}
\includegraphics[scale=0.7]{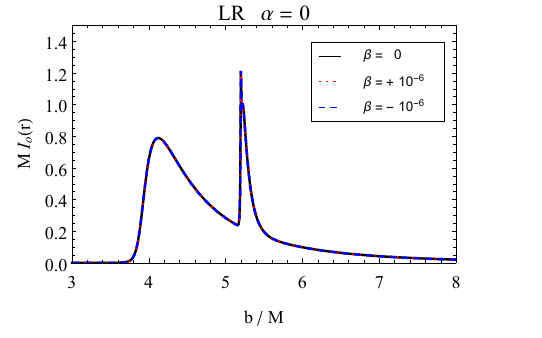}
\includegraphics[scale=0.7]{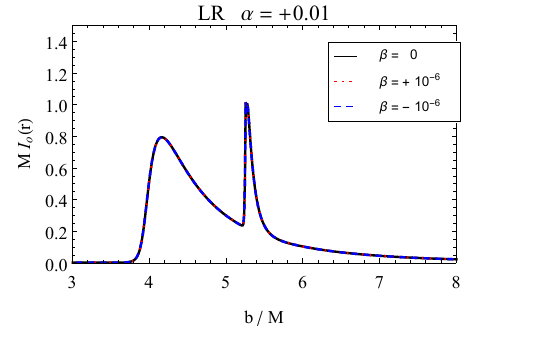}
\caption{Intensity $I(r)$ as a function of the radius $r$ for the LR disk model for a varying $\alpha$ with constant $\beta$ (top row) and for a varying $\beta$ with constant $\alpha$ (bottom row)}
\label{fig:intensity_LR_A}
\end{figure*}

\begin{figure*}
    \centering
    \includegraphics[scale=0.5]{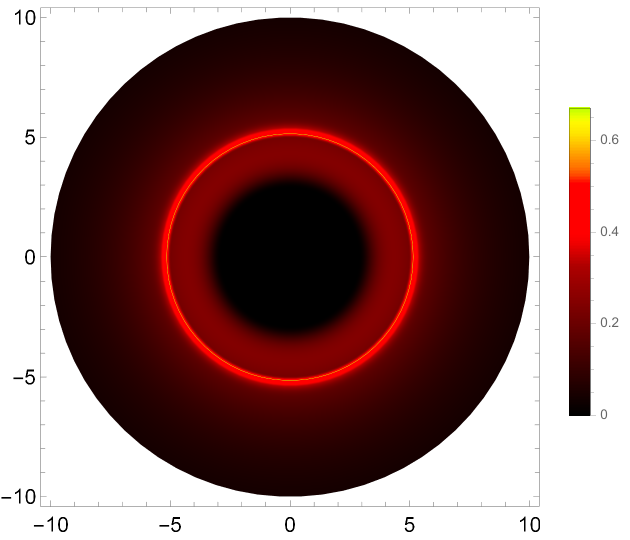}
    \includegraphics[scale=0.5]{plot_am3_b0_EH.pdf}
    \includegraphics[scale=0.5]{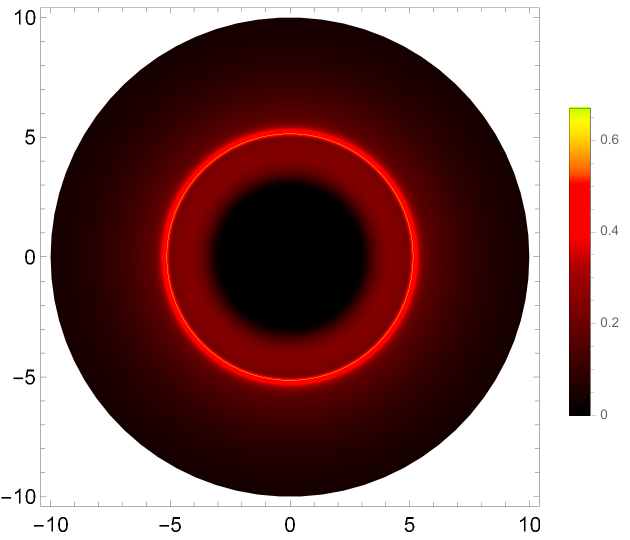}\\
    \includegraphics[scale=0.5]{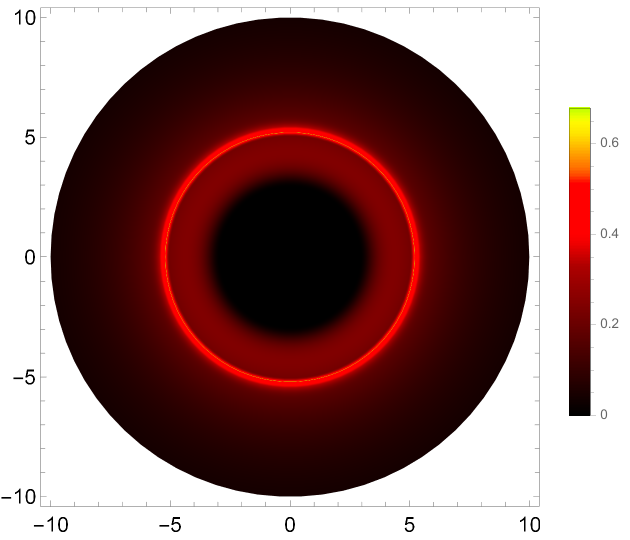}
    \includegraphics[scale=0.5]{plot_a0_b0_EH.pdf}
    \includegraphics[scale=0.5]{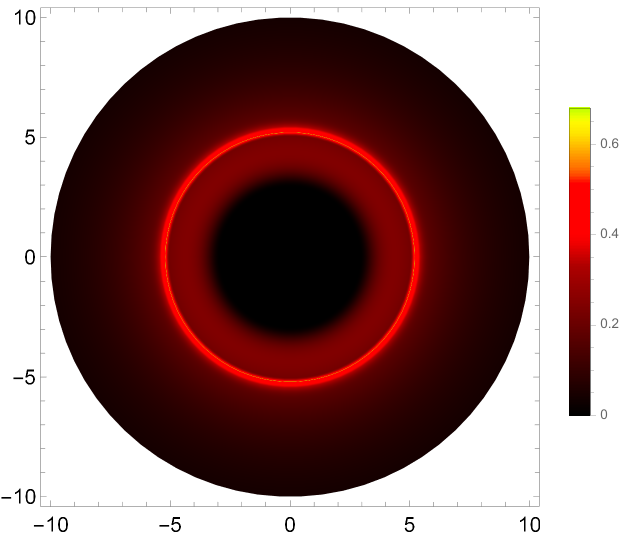}\\
    \includegraphics[scale=0.5]{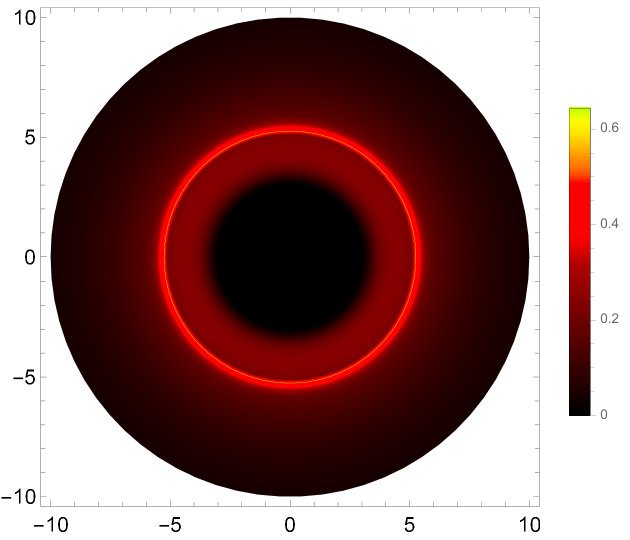}
    \includegraphics[scale=0.5]{plot_ap3_b0_EH.pdf}
    \includegraphics[scale=0.5]{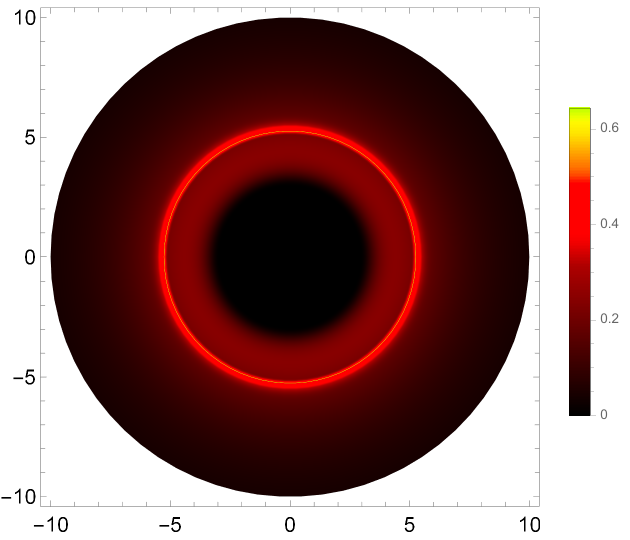}\\
    \caption{Shadow images for the EH disk model with different combinations of $\alpha$ and $\beta$, namely $\alpha=-0.01$ (top row), $\alpha=0$ (middle row), and $\alpha=+0.01$ (bottom row); and $\beta=-10^{-5}$ (left column), $\beta=0$ (middle column), and $\beta=10^{-5}$ (right) column). }
    \label{fig:shadows_EH_A}
\end{figure*}

\begin{figure*}
\includegraphics[scale=0.7]{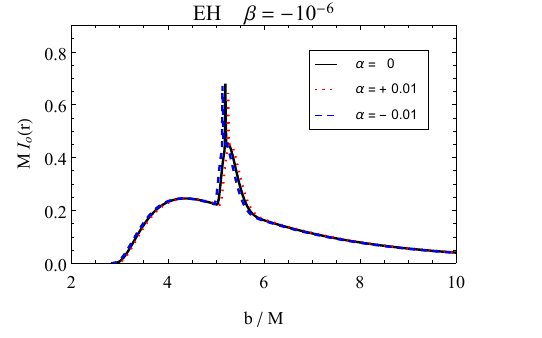}
\includegraphics[scale=0.7]{int_b0_EH.pdf}
\includegraphics[scale=0.7]{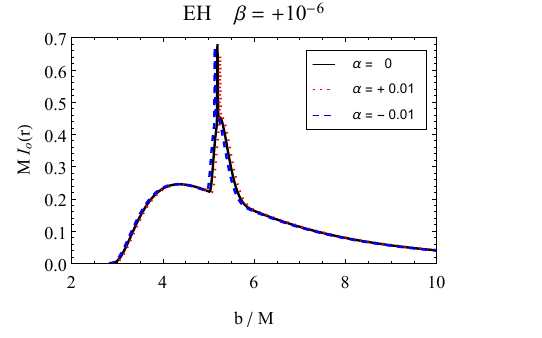}
\includegraphics[scale=0.7]{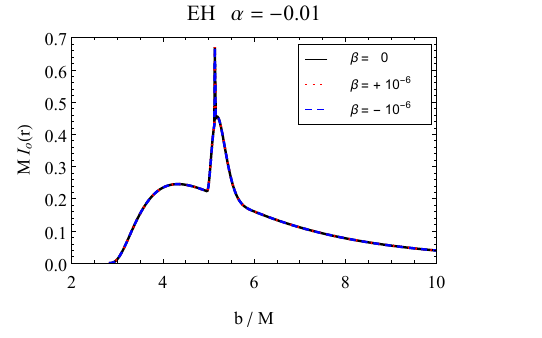}
\includegraphics[scale=0.7]{int_a0_EH.pdf}
\includegraphics[scale=0.7]{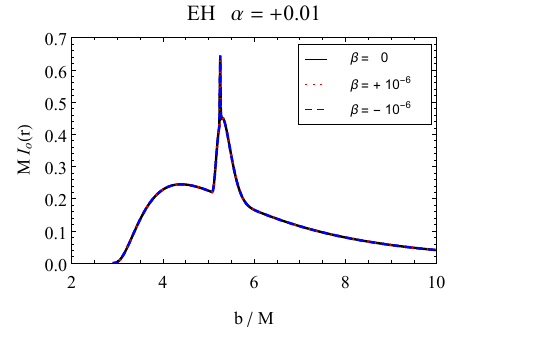}
\caption{Intensity $I(r)$ as a function of the radius $r$ for the EH disk model for a varying $\alpha$ with constant $\beta$ (top row) and for a varying $\beta$ with constant $\alpha$ (bottom row).}
\label{fig:intensity_EH_A}
\end{figure*}

For completeness, in this appendix we show the complete set of shadow images and intensity plots for all of the parameter combinations, i.e., $\alpha=\{-0.01, 0, 0.01\}$ and $\beta=\{-10^{-5}, 0, 10^{-5}\}$, and for all of the accretion disk models, i.e., the ISCO, the LR, and the EH model in Figures \ref{fig:shadows_ISCO_A}-\ref{fig:intensity_EH_A}. These results serve to illustrate the negligible effect of the parameter $\beta$ to the observable properties of the system.

\end{appendix}

\end{document}